\documentclass[a4paper,fleqn,usenatbib]{mnras}

\pdfoutput=1

\usepackage{newtxtext,newtxmath}

\usepackage[utf8]{inputenc}
\usepackage[T1]{fontenc}
\usepackage{ae,aecompl}

\usepackage{graphicx}	
\usepackage{amsmath}	
\usepackage{amssymb}	
\usepackage{color}
\usepackage{xspace}

\newcommand{\etal}{et\thinspace al.\thinspace}
\newcommand{\starlight}{\textsc{starlight}\xspace}
\newcommand{\pycasso}{PyCASSO\xspace}
\newcommand{\sfourg}{S\textsuperscript{4}G\xspace}

\title[The PyCASSO database]{The PyCASSO database: Spatially resolved stellar
population properties for CALIFA galaxies}

\author[A. L.\ de Amorim et al.]{A. L. de Amorim,$^{1}$\thanks{E-mail: andre@astro.ufsc.br}
R. Garc\'{\i}a-Benito,$^{2}$ 
R. Cid Fernandes,$^{1}$
C. Cortijo-Ferrero,$^{2}$
\newauthor
R. M. Gonz\'alez Delgado,$^{2}$
E. A. D. Lacerda,$^{1}$
R. L\'opez Fern\'andez,$^{2}$
E. P\'erez,$^{2}$
\newauthor
and N. Vale Asari$^{1}$
\\
$^{1}$Departamento de F\'{\i}sica, Universidade Federal de Santa
Catarina, P.O. Box 476, 88040-900, Florian\'opolis, SC, Brazil \\
$^{2}$Instituto de Astrof\'{\i}sica de Andaluc\'{\i}a
(CSIC), P.O. Box 3004, 18080 Granada, Spain.
}

\date{Accepted XXX. Received YYY; in original form ZZZ}

\pubyear{2017}

\begin{document}
\label{firstpage}
\pagerange{\pageref{firstpage}--\pageref{lastpage}}
\maketitle
   
\begin{abstract}
The Calar Alto Legacy Integral Field Area (CALIFA) survey, a pioneer in integral
field spectroscopy legacy projects, has fostered many studies exploring the
information encoded on the spatially resolved data on gaseous and stellar
features in the optical range of galaxies.
We describe a value-added catalogue of stellar population properties for CALIFA
galaxies analysed with the spectral synthesis code \starlight and processed with
the \pycasso platform. Our public data base (\texttt{http://pycasso.ufsc.br/},
mirror at \texttt{http://pycasso.iaa.es/}) comprises 445 galaxies from the
CALIFA Data Release 3 with COMBO data.
The catalogue provides maps for the stellar mass surface density, mean stellar
ages and metallicities, stellar dust attenuation,  star formation rates, and
kinematics. Example applications both for individual galaxies and for
statistical studies are presented to illustrate the power of this data set.
We revisit and update a few of our own results on mass density radial profiles
and on the local mass--metallicity relation.
We also show how to employ the catalogue for new investigations, and show a
pseudo Schmidt-Kennicutt relation entirely made with information extracted from
the stellar continuum.
Combinations to other databases are also illustrated. Among other
results, we find a very good agreement between star formation rate surface
densities derived from the stellar continuum and the $\mathrm{H}\alpha$
emission.
This public catalogue joins the scientific community's effort towards
transparency and reproducibility, and will be useful for researchers focusing on
(or complementing their studies with) stellar properties of CALIFA galaxies.
\end{abstract}


\section{Introduction}
\label{sec:intro}
The {\em modus operandi} of extragalactic astrophysics was transformed by large
spectroscopic surveys, such as the 2dF Galaxy Redshift Survey
\citep{Colless.etal.2001a} and the Sloan Digital Sky Survey
\citep[SDSS;][]{York.etal.2000a}. Instead of relying on a few dozen galaxies,
now hundreds of thousands are routinely summoned to address a wide variety of
astrophysical questions. The combination of large surveys with stellar
population libraries with good spectral resolution
\citep[e.g.][]{Bruzual.Charlot.2003a} was the cornerstone for the uprise of
several spectral synthesis codes, such as our own code \starlight
\citep{CidFernandes.etal.2005a}, and those by other groups: {\scshape moped}
\citep{Panter.Heavens.Jimenez.2003a}, {\scshape stecmap/steckmap}
\citep{Ocvirk.etal.2006a,
  Ocvirk.etal.2006b}, {\scshape vespa} \citep{Tojeiro.etal.2007a},
{\scshape ulyss} \citep{Koleva.etal.2008a}.

One example of the statistical power of applying \starlight to a million SDSS
galaxy spectra was revealing a huge and forgotten population of retired galaxies
ionized by hot low mass evolved stars, whose spectra are confounded with those
of low ionization nuclear emission regions \citep{Stasinska.etal.2008a,
CidFernandes.etal.2010a, CidFernandes.etal.2011a}. This population completely
changes the census of galaxies in the local universe
\citep{Stasinska.etal.2015b}.
Other examples include studies of the stellar mass-metallicity relation
\citep{ValeAsari.etal.2009a}, the chemical evolution and the star formation
history of local galaxies \citep{CidFernandes.etal.2007a,Asari.etal.2007a}, how
to distinguish AGN hosts \citep{Stasinska.etal.2006a}, the bimodality of
galaxies and the downsizing \citep{Mateus.etal.2006a}, and the dependence of
galaxy properties on the environment \citep{Mateus.etal.2007a}.  More
importantly in the context of this paper, the data and tools behind these
studies were made publicly available, fostering independent research. Since
having published our value-added \starlight-SDSS catalogue
\citep{CidFernandes.etal.2009a} in a \emph{public} database
(\url{http://starlight.ufsc.br}), studies like \citet{Bian.etal.2007a,
  Liang.etal.2007a, Peeples.Pogge.Stanek.2009a, Riffel.etal.2009b,
  LaraLopez.etal.2009a, LaraLopez.etal.2009b, LaraLopez.etal.2010a,
  Andrews.Martini.2013a} have used our code and database.

These past few years have witnessed the surge of integral field spectroscopy
(IFS) observations. While SDSS obtained one spectrum per galaxy, IFS surveys
commit to spectrally map galaxies pixel by pixel.
The Calar Alto Legacy Integral Field Survey Area survey
\citep[CALIFA;][]{Sanchez.etal.2012a} was a pioneer wide-field IFS survey of
nearby galaxies. Its Data Release 3 \citep[DR3;][]{Sanchez.etal.2016c} contains
445 galaxy COMBO data cubes covering $3700$--$7500$ \AA\ and in the redshift
range $0.005 < z < 0.03$. The COMBO cubes have a resolution of $850$,
approximately $\mathrm{FWHM} = 6\,\text{\AA}$ at $\lambda = 5000\,\text{\AA}$.
The spatial sampling is $1^{\prime\prime}$, with a PSF given by a Moffat profile
with $\mathrm{FWHM} = 2.5^{\prime\prime}$ and $\beta=2$. CALIFA provides a
statistically significant sample which is complete in the magnitude range $-19.0
< M_r < -23.1$ and stellar masses\footnote{Assuming a Salpeter IMF.} between
$10^{9.7}$ and $10^{11.4}\,\mathrm{M}_\odot$ \citep{Walcher.etal.2014a}. In
addition to its clear selection criteria, another characteristic that sets
CALIFA apart is its spatial coverage ($> 2$ half light radius), which maps
galaxies farther in their outskirts than other ongoing surveys such as SAMI
\citep{Croom.etal.2012a} and MaNGA \citep{Bundy.etal.2015a}.

We have analysed the CALIFA data cubes with our spectral synthesis code
\starlight and the Python CALIFA \starlight Synthesis organiser (\pycasso)
platform, as described in \citet{CidFernandes.etal.2013a,
CidFernandes.etal.2014a}.This formed the basis of a series of studies about the
spatially resolved stellar population properties, where we have obtained:

\begin{enumerate}
\item
Evolutionary curves of the cumulative mass function, tracing the mass
assembly history of $\sim 100$ galaxies as a function of the radial
distance \citep{Perez.etal.2013a}. 
The results  suggest that galaxies grow inside-out, as  confirmed by 
García-Benito et al. (in prep.) for a seven times larger sample.

\item 2D maps and spatially resolved information of the stellar populations
properties, used to retrieve the local relations between: a) stellar mass
surface density, $\Sigma_\star$, and luminosity weighted stellar ages, $\langle
\log\,t \rangle_\mathrm{L}$ \citep{GonzalezDelgado.etal.2014b}; b) mass weighted
stellar metallicity, $\langle \log\,Z \rangle_\mathrm{M}$, and $\Sigma_\star$
\citep{GonzalezDelgado.etal.2014a}; c) the intensity of the star formation rate,
$\Sigma_\mathrm{SFR}$, defined in this work as the star formation rate surface
density, and $\Sigma_\star$ \citep{GonzalezDelgado.etal.2016a}. These relations
indicate that local processes are relevant to regulate the star formation and
chemical enrichment in the disk of spirals. For spheroids, on the other hand,
the stellar mass, $M_\star$, regulates these processes.

\item Radial profiles of stellar extinction, $A_V$,  $\Sigma_\star$, $\langle
\log\,t \rangle_\mathrm{L}$, $\langle \log\,Z \rangle_\mathrm{M}$, and their
radial gradients also confirm that more massive galaxies are more compact,
older, metal rich, and less reddened by dust. These trends are preserved
spatially as a function of the radial distance to the nucleus.
Deviations from these relations seem to be correlated with Hubble type: earlier
types are more compact, older, and more metal rich for a given $M_\star$, which
evidences that the shut down of the star formation is related to galaxy
morphology \citep{GonzalezDelgado.etal.2015c}. The negative radial gradients of
$\langle \log\,t \rangle_\mathrm{L}$ and $\langle \log\,Z \rangle_\mathrm{M}$
also confirm that galaxies grow inside-out.

\item The radial profiles of $\Sigma_\mathrm{SFR}$ and the small dispersion
between the profiles of late type spirals (Sbc, Sc, Sd) confirm that the main
sequence of the star forming galaxies is a sequence with nearly constant
$\Sigma_\mathrm{SFR}$ \citep{GonzalezDelgado.etal.2016a}. Furthermore, the
radial profiles of local and recent specific star formation rate depend on
Hubble type (increasing as one goes from early to late types) and increase
radially outwards. This behavior suggests that galaxies are quenched inside-out,
and that this process is faster in the central bulge-dominated parts than in the
disks \citep{GonzalezDelgado.etal.2016a}.

\item From the 2D ($R\,\times\,t$) map of the galaxy SFH we retrieve the
spatially resolved time evolution information of the star formation rate (SFR),
its intensity ($\Sigma_\mathrm{SFR}$), and specific star formation (sSFR).
We find that galaxies form very fast independently of their stellar mass, with
the peak of star formation at high redshift ($z > 2$), and that subsequent star
formation is driven by $M_\star$ and morphology, with less massive and later
type spirals showing more prolonged periods of star formation (González Delgado
\etal 2017, submitted).

\item From the comparison of the spatially resolved and the integrated spectra
analyses, we find that the stellar population properties are well represented by
their values at one half light radius \citep{GonzalezDelgado.etal.2014b}.

\item Detailed studies of a small sample of galaxies are very well served by the
CALIFA survey and our analysis. For example, the analysis of the spatially
resolved SFH of mergers (\citealt{CortijoFerrero.etal.2017a}; Cortijo-Ferrero et
al. 2017b, 2017c, in prep.) in comparison with ``non-interacting'' spiral
galaxies allows us to estimate the effect of the merger phase in the enhancement
of the star formation and their spatial extent, as well as to trace the merger
epochs.

\item By extending \starlight to combine CALIFA optical spectra and GALEX
\citep{Martin.etal.2005a} UV photometric data, the uncertainties in stellar
properties become smaller \citep{LopezFernandez.etal.2016a}. Also, young stellar
populations are better constrained, especially for low mass late-type galaxies,
where there is a significant $<300\,\mathrm{Myr}$ population.

\item  The well defined selection function of the survey allows for reliable
volume corrections \citep{Walcher.etal.2014a}. Combining these with our
homogeneous analysis led to an estimation of the SFR density of the Local
Universe of $0.0105 \pm 0.0008 \,$M$_\odot\,$yr$^{-1}\,$Mpc$^{-3}$, in agreement
with independent estimates (\citealt{GonzalezDelgado.etal.2016a}; see also López
Fernández et al., in prep.).

\end{enumerate}

We release the data used in those papers in a public value-added catalogue
(\url{http://pycasso.ufsc.br/}, mirror at \url{http://pycasso.iaa.es/}).
The data were updated to the latest CALIFA release (DR3) and reduction pipeline
(v2.2).  Stellar population models used in the \starlight fits were also
updated. These modifications have only minor effects with respect to the results
reported in the studies mentioned above. Still, when exemplifying the use of
this new database we will take the opportunity to revisit and update some of our
previous results.

In addition to embracing the ethos of open science data and tools to encourage
reproducible results, our CALIFA-\starlight value-added catalogue can be used
for many different purposes. We list a few of those, based on the impact of our
former SDSS-\starlight public database: (a) Stellar population data may be used
to complement studies focused on nebular properties. (b) Our catalogue can be
used to investigate the host galaxy of transient sources. For instance, if there
is a supernova explosion in a CALIFA galaxy ten years from now, interested
researchers need simply to look up the corresponding spaxel in our catalogue
\citep{Stanishev.etal.2012a, Galbany.etal.2014a}. (c) Finally, other groups
might be interested in the stellar population themselves, using our database
with a different perspective from our group.

This paper is organised as follows. Section \ref{sec:obs-sample} describes the
CALIFA observations and the sample selection. In section \ref{sec:analysis} we
discuss how CALIFA galaxies are analysed with \starlight, our spectral synthesis
code, and \pycasso, a tool to organise the synthesis results for IFS data.
Section \ref{sec:catalogue} describes the physical and quality control maps
available in our value-added catalogue, as well as details on how to access the
data. Section \ref{sec:using-the-catalogue} shows a few example usages for the
catalogue. Some comparisons to data from other sources are presented in Section
\ref{sec:example-other}. Our concluding remarks follow in Section
\ref{sec:conclusions}.

\section{Observations and sample}
\label{sec:obs-sample}
The survey data was collected using the Potsdam Multi-Aperture Spectrophotometer
\citep[PMAS]{Roth.etal.2005a}, in PPak mode \citep{Kelz.etal.2006a}. PPak is a
bundle of 382 fibers, each one having $2.7^{\prime\prime}$ of diameter. It
covers a field of view (FoV) of $74^{\prime\prime} \times 64^{\prime\prime}$,
with a filling factor of $\sim60\%$.

Observations for CALIFA were obtained in two different spectral settings using
the V500 and V1200 gratings. The V500 grating has a spectral resolution of
$\sim\,6\,\text{\AA}$ (FWHM), with a wavelength coverage from
$3745$--$7300\,\text{\AA}$, while the V1200 has a higher spectral resolution of
$\sim\,2.3\,\text{\AA}$, covering the $3650$--$4840\,\text{\AA}$ range. However,
vignetting on the CCD corners reduce the useful wavelength range in some regions
of the FoV. The vignetting can be reduced by combining observations using V500
and V1200 gratings. These are called COMBO cubes. A dithering scheme using three
observations is used to fill the whole FoV. More information on the
observational strategy, effects of vignetting, the reduction pipeline, and data
quality can be found in \citet{Sanchez.etal.2012a} and
\citet{Sanchez.etal.2016c}.

The sample is formed by 445 galaxies of the final CALIFA data release DR3
\citep{Sanchez.etal.2016c} that were observed with both V500 and V1200, and have
COMBO cubes. This sample includes 395 galaxies from the main sample, the
remaining galaxies are from the extended CALIFA sample. The main sample is a
statistically significant subset of the mother sample, which in turn was drawn
from the SDSS DR7 \citep{Abazajian.etal.2009a}. It consists of a representative
sample of galaxies of the local universe at the redshift range $0.005 < z <
0.03$, with magnitudes ($-24 < M_r < -17$) and colors ($u-r < 5$) that cover all
the color magnitude diagram (CMD). The galaxies were selected by their apparent
size with an angular isophotal diameter $45^{\prime\prime} < \mathrm{isoA}_r <
80^{\prime\prime}$ to fill the FoV of PPaK. A full description of the mother
sample is in \citet{Walcher.etal.2014a}. The extended CALIFA sample comes from
an heterogeneous set of galaxies observed in different ancillary science
projects that are fully described in \citet{Sanchez.etal.2016c}.

\begin{figure}
\centering
\includegraphics{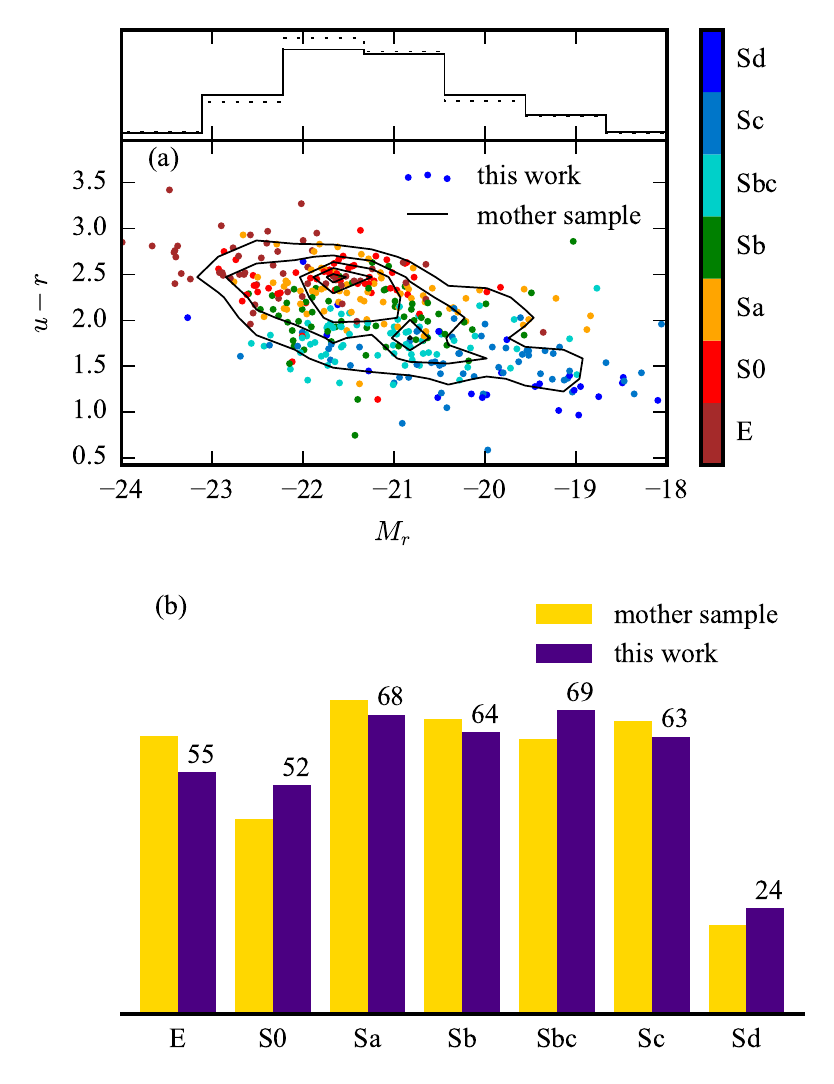}
\caption{{\em Panel (a)} Color--magnitude diagram (CMD) of the sample, using
photometric magnitudes from SDSS DR7. The black contours show the density of
galaxies in the mother sample. The dots, colored by morphological type, show the
395 galaxies with COMBO cubes from the main sample. Above the CMD is the $M_r$
histogram of both the mother sample, as a black line, and the 395 galaxy
sample, as a dotted line. {\em Panel (b)} Histogram of morphological types of
the mother sample (red bars) and the 395 galaxy sample (blue bars).
The histograms are normalized for comparison between the two distributions. The
number of galaxies in each type for the 395 galaxies from the main sample is
printed above the blue bars.}
\label{fig:cmd-morph}
\end{figure}

The galaxies from the main and the extended samples were morphologically
classified by members of the collaboration through a visual inspection of the
SDSS r-band images\footnote{Morphological classification performed by the
collaboration is available as ancillary data tables, at the CALIFA DR3 website
(\url{http://www.caha.es/CALIFA/public_html/?q=content/dr3-tables}).}.
Fig.\ \ref{fig:cmd-morph}a shows the location of the main sample galaxies in the
CMD, compared to the mother sample. The distribution by morphological type,
again for both samples, is shown in Fig.\ \ref{fig:cmd-morph}b.

\section{Method of analysis}
\label{sec:analysis}
We use a full spectral approach to the stellar population synthesis, using
\starlight. The cubes pass through a series of pre-processing steps, meant to
extract good quality data for the analysis. These are part of the QBICK
pipeline, described in detail in the section 3.2 of
\citet{CidFernandes.etal.2013a}. The main steps are briefly discussed below.

\subsection{Pre-processing}
\label{sec:preprocessing}
The first step is to mask all spaxels containing light from spurious sources.
That includes foreground stars and other galaxies in the FoV. The outer spaxels
that have $\mathrm{S}/\mathrm{N} < 3$ are also masked.

The cubes are then segmented in Voronoi zones, as implemented by
\citet{Cappellari.Copin.2003a}. We use, as input, signal and noise images from
the $5635 \pm 45\,\text{\AA}$ spectral range (the same that is used for spectra
normalization, as discussed below). Some modifications were done on the coadded
error estimation to account for spatial covariance, as described in the appendix
of \citet{GarciaBenito.etal.2015a}. The target S/N was $20$, which lead to most
zones inside one half light radius consisting of a single spaxel.

\subsection{Spectral synthesis}
\label{sec:spec-synthesis}

After the pre-processing steps the resulting spectra are fitted using \starlight
\citep{CidFernandes.etal.2005a}. We fit the  $\lambda = 3700$ to
$6850\,\text{\AA}$ rest-frame stellar continuum, masking spectral windows around
emission lines ([O~\textsc{ii}]$\lambda 3726/3729$, H$\gamma$, H$\beta$,
[O~\textsc{iii}]$\lambda 5007$, He~\textsc{i}$\lambda 5876$,
[O~\textsc{i}]$\lambda 6300$, H$\alpha$, [N~\textsc{ii}]$\lambda 6584$,
[S~\textsc{ii}]$\lambda 6716/6731$). We further mask the Na~\textsc{i} doublet
at $\lambda = 5890\,\text{\AA}$ because of possible absorption in the neutral
interstellar medium.

The fits decompose the observed spectra in a base built from single stellar
populations (SSP). Model SSP spectra are a central ingredient in our analysis,
that allow us to link the results of the spectral decomposition to physical
properties of the stellar populations. Our previous analysis of CALIFA data
(reviewed in section \ref{sec:intro}) explored different sets of SSP spectra,
varying the number of components, the range in metallicity, the specific ages of
the SSP, initial mass function (IMF), and stellar evolution.  The fits presented
here were obtained using two sets of bases, labeled as \textit{GMe} and
\textit{CBe}, that have been extensively tested and compared
(\citealt{GonzalezDelgado.etal.2015a}, \citealt{GonzalezDelgado.etal.2015b},
\citealt{GonzalezDelgado.etal.2016a}). These bases are similar to base
\textit{GM} and \textit{CB} of \citet{CidFernandes.etal.2014a}, but extended in
terms of the metallicity coverage.

Base \textit{GMe} is a combination of 235 SSP spectra from
\citet{Vazdekis.etal.2010a}, for populations older than $63\,\mathrm{Myr}$, and
\citet{GonzalezDelgado.etal.2005a} models, for younger ages. The evolutionary
tracks are those from \citet{Girardi.etal.2000a}, except for the youngest ages
($1$ and $3\,\mathrm{Myr}$), which are based on the Geneva tracks
\citep{Schaller.etal.1992a, Schaerer.etal.1993a, Charbonnel.etal.1993a,
Schaerer.etal.1993b}. The metallicity covers the seven bins provided by
\citep{Vazdekis.etal.2010a}: $\log\,Z/\mathrm{Z}_\odot = -2.3, -1.7, -1.3, -0.7,
-0.4, 0, +0.22$, but for SSP younger than $63\,\mathrm{Myr}$, $Z$ includes only
the the four largest metallicities. The chosen IMF is Salpeter.

Base \textit{CBe} is a combination of 246 SSP spectra provided by Charlot \&
Bruzual (2007, private communication\footnote{The Charlot \& Bruzual (2007)
models are available at \url{http://www.bruzual.org/~gbruzual/cb07}.}), covering
41 ages from $0.001$ to $14\,\mathrm{Gyr}$, with six metallicities: $\log \
Z/\mathrm{Z}_\odot = -2.3, -1.7, -0.7, -0.4, 0, +0.4$. The IMF is that from
\citet{Chabrier.2003a}. These SSP are an update of the
\citet{Bruzual.Charlot.2003a} models, where STELIB \citep{LeBorgne.etal.2003a}
is replaced by the MILES
\citep{SanchezBlazquez.etal.2006a} and GRANADA \citep{martins.etal.2005a}
spectral libraries. The stellar tracks are those known as ``Padova 1994''
\citep{Alongi.etal.1993a, Bressan.etal.1993a, Fagotto.etal.1994a,
Fagotto.etal.1994b, Fagotto.etal.1994c, Girardi.etal.1996a}.

Other important issues in our methodology are: a) dust effects are also modeled
in the spectral fits as a foreground screen with a
\citet*{Cardelli.Clayton.Mathis.1989a} reddening law with $R_V = 3.1$; b)
kinematics effects are also taken into account, assuming a Gaussian
line-of-sight velocity distribution. Both spectral bases are smoothed to
$6\,\text{\AA}$ FWHM effective resolution prior to the fits. This is because the
kinematical filter implemented in \starlight operates in velocity-space, whereas
both CALIFA and the SSP spectra have a constant resolution in $\lambda$-space.
This way the velocity dispersion obtained by \starlight is not contaminated by
the spectral resolution mismatch.

\subsection{PyCASSO}
\starlight was developed to work with individual spectra. The \pycasso\ platform
was developed to organise its output into spatially resolved data.

The output produced by \starlight is quite large, and it takes a little effort
to transform some of the values into familiar physical measurements. \pycasso
takes care of these computations and provides convenient functions that allow us
to deal directly with the physical maps. This made the exploration of the data a
lot easier, yielding the various results discussed in section \ref{sec:intro}.
These physical maps (and some diagnostic ones) are published in this database,
and are described in the following section.

\section{Description of the catalogue}
\label{sec:catalogue}
The catalogue is comprised of FITS files containing images for a suite of physical
properties of each galaxy of the sample. The files also contain quality control
and fit diagnostic maps. A FITS table with properties derived for the integrated
spectra of the galaxies is also available. We present these data products below.

\subsection{Map data format}
\label{sec:data-format}
Maps are distributed in multi-extension binary FITS files, containing several
HDUs. The HDUs consist of a header and an image extension. The headers are all
the same for all HDUs, except for the first (primary), which contains additional
cards. In particular, \texttt{CALIFA\_ID} and \texttt{NED\_NAME} are present for
object identification. \texttt{BASE} informs which SSP base was used in the fit;
it may be either \texttt{gsd6e} (the \textit{GMe} base) or \texttt{zca6e} (the
\textit{CBe} base). Also worth mentioning is the luminosity distance used when
calculating masses, stored in the card \texttt{DL\_MPC}, in units of Mpc. All
headers also contain celestial WCS extracted from the original COMBO cubes.

\begin{table*}
	\caption{Physical quantities and diagnostic maps present in the galaxy cubes.}
	\label{tab:maps}
	\centering
	\begin{tabular}{r l l l l}
		\hline\hline
		HDU & Quantity & Units & Extension name & Description \\
		\hline
		0 & $\Sigma_\star$ & $\mathrm{M}_\odot\,\mathrm{pc}^{-2}$ &
		\texttt{sigma\_star} & Stellar mass surface density. Mass that is currently
		trapped inside stars.\\
		1 & $\Sigma^\prime_\star$ & $\mathrm{M}_\odot\,\mathrm{pc}^{-2}$ &
		\texttt{sigma\_star\_ini} & Initial stellar mass surface density. Mass
		ever formed into stars.\\
		2 & $\mathcal{L}_{5635\text{\AA}}$ &
		$\mathrm{L}_\odot\,\text{\AA}^{-1}\,\mathrm{pc}^{-2}$ & \texttt{L\_5635} &
		Luminosity surface density in the $5635$ \AA\ normalization window.\\
		3 & $\langle \log\,t \rangle_\mathrm{L}$ & $\log\,\mathrm{yr}$ &
		\texttt{log\_age\_flux} & Mean log of stellar age, in years, weighted by
		luminosity.\\
		4 & $\langle \log\,t \rangle_\mathrm{M}$ & $\log\,\mathrm{yr}$ &
		\texttt{log\_age\_mass} & Mean log of stellar age, in years, weighted by
		mass.\\
		5 & $\langle \log\,Z \rangle_\mathrm{L}$ & $\log\,\mathrm{Z}_\odot$ &
		\texttt{log\_Z\_flux} & Mean log of stellar metallicity, in solar units,
		weighted by luminosity.\\
		6 & $\langle \log\,Z \rangle_\mathrm{M}$ & $\log\,\mathrm{Z}_\odot$ &
		\texttt{log\_Z\_mass} & Mean log of stellar metallicity, in solar units,
		weighted by mass.\\
		7 & $\Sigma_\mathrm{SFR}$ &
		$\mathrm{M}_\odot\,\mathrm{Gyr}^{-1}\,\mathrm{pc}^{-2}$ & \texttt{sigma\_sfr}
		& Mean star formation rate surface density in the last $32\,\mathrm{Myr}$.\\
		8 & $x_\mathrm{Y}$ & - & \texttt{x\_young}
		& Luminosity fraction of stellar populations younger than
		$32\,\mathrm{Myr}$.\\
		9 & $\tau_V$ & - &
		\texttt{tau\_V} & Attenuation coefficient for a dust screen model.\\
		10 & $v_\star$ & $\mathrm{km\,s}^{-1}$ &
		\texttt{v\_0} & Line of sight velocity.\\
		11 & $\sigma_\star$ & $\mathrm{km\,s}^{-1}$ &
		\texttt{v\_d} & Line of sight velocity dispersion.\\
		\\
		12 & $\overline\Delta$ & \% & \texttt{adev} & Mean model deviation.\\
		13 & $N_{\lambda,\mathrm{clip}} / N_\lambda$ & \% & \texttt{nl\_clip} &
		Fraction of the wavelengths clipped by the fitting algorithm.\\
		14 & $\chi^2/N_\lambda^\mathrm{eff}$ & - & \texttt{chi2} & Fit statistic.\\
		15 & Bad pixels & - & \texttt{badpix} & Masked pixels. \\
		16 & Zones & - & \texttt{zones} & Voronoi segmentation zones.\\
		17 & S/N & - & \texttt{sn} & Signal--to--noise ratio in individual pixels.\\
		18 & S/N (zones) & - & \texttt{sn\_zone} & Signal--to--noise ratio in zones.\\
		\hline
	\end{tabular}
\end{table*}

For a given galaxy, the maps are images with the same dimensions. The exact
dimensions will change for each galaxy; they are the same as the raw cubes from
DR3. Except for \texttt{badpix} and \texttt{zones}, which are integers, all data
are stored as 64-bit floating point numbers. If a pixel is considered ``bad''
(see Section \ref{sec:data-quality}), its value in all maps is set to
zero\footnote{The bad pixels values are set to zero, which may be a valid value
depending on the map. A map indicating which pixels are bad is also provided, to
help in those cases.}.
The 12 first HDUs, listed in Table \ref{tab:maps}, contain maps with physical
results from \starlight. The remaining 7 HDUs contain data and fit diagnostic
maps. The content of these maps are explored in detail in Sections
\ref{sec:physical-maps} and \ref{sec:quality-control}.

\subsection{Integrated spectra table data format}
We also fit the integrated spectra of the galaxies, summing the flux from all
non-masked spaxels. The data are stored in a binary FITS table, with columns
described in Table \ref{tab:integrated}. The data are stored as 64-bit floating
point numbers, except for \texttt{califaID}, which is integer, and
\texttt{NED\_name} and \texttt{base}, which are text strings. The quantities in
this table are the same ones provided as maps. Note, however, that the extensive
quantities in this table (luminosity, stellar masses, and SFR) are not surface
densities.

\begin{table*}
	\caption{Global properties of galaxies using integrated spectra.}
	\label{tab:integrated}
	\centering
	\begin{tabular}{l l l l}
		\hline\hline
		Quantity & Units & Column & Description \\
		\hline
		CALIFA ID & - & \texttt{califaID} & CALIFA galaxy ID.\\
		Object name & - & \texttt{NED\_name} & Object name.\\
		Luminosity distance & Mpc & \texttt{DL\_Mpc} & Luminosity distance used
		when calculating stellar masses.\\
		SSP base & - & \texttt{base} & Base used in the fit.\\
		\\
		$M_\star$ & $\mathrm{M}_\odot$ &
		\texttt{M\_star} & Stellar mass currently trapped inside stars.\\
		$M^\prime_\star$ & $\mathrm{M}_\odot$ &
		\texttt{M\_star\_ini} & Total mass ever formed into stars.\\
		$L_{5635\text{\AA}}$ & $\mathrm{L}_\odot\ \text{\AA}^{-1}$
		& \texttt{L\_5635} & Luminosity in the $5635$ \AA\ normalization
		window.\\
		$\langle \log\,t \rangle_\mathrm{L}$ & $\log\ \mathrm{yr}$ &
		\texttt{log\_age\_flux} & Mean log of stellar age, in years, weighted by
		luminosity.\\
		$\langle \log\,t \rangle_\mathrm{M}$ & $\log\ \mathrm{yr}$ &
		\texttt{log\_age\_mass} & Mean log of stellar age, in years, weighted by
		mass.\\
		$\langle \log\,Z \rangle_\mathrm{L}$ & $\log\ \mathrm{Z}_\odot$ &
		\texttt{log\_Z\_flux} & Mean log of stellar metallicity, in solar units,
		weighted by luminosity.\\
		$\langle \log\,Z \rangle_\mathrm{M}$ & $\log\ \mathrm{Z}_\odot$ &
		\texttt{log\_Z\_mass} & Mean log of stellar metallicity, in solar units,
		weighted by mass.\\
		$\mathrm{SFR}$ & $\mathrm{M}_\odot\ \mathrm{Gyr}^{-1}$ & \texttt{sfr} &
		Mean star formation rate in the last $32\,\mathrm{Myr}$.\\
		$x_\mathrm{Y}$ & - & \texttt{x\_young} & Luminosity fraction of stellar
		populations younger than $32\,\mathrm{Myr}$.\\
		$\tau_V$ & - &
		\texttt{tau\_V} & Attenuation coefficient for a dust screen model.\\
		$v_\star$ & $\mathrm{km\ s}^{-1}$ &
		\texttt{v\_0} & Line of sight velocity.\\
		$\sigma_\star$ & $\mathrm{km\ s}^{-1}$ &
		\texttt{v\_d} & Line of sight velocity dispersion.\\
		\\
		$\chi^2/N_\lambda^\mathrm{eff}$ & - & \texttt{chi2} & Fit statistic.\\
		$\overline\Delta$ & \% & \texttt{adev} & Mean model deviation.\\
		$N_{\lambda,\mathrm{clip}} / N_\lambda$ & \% & \texttt{nl\_clip} & Fraction
		of the wavelengths clipped by the fitting algorithm.\\
		
		\hline
	\end{tabular}
\end{table*}

\subsection{Data access}
\label{sec:data-access}

The data are publicly available at the \pycasso web site
(\url{http://pycasso.ufsc.br/}, mirror at \url{http://pycasso.iaa.es/}). The
galaxies are listed with a color SDSS image stamp for reference. There are some
preview plots available for each galaxy, in the same fashion as Fig.\
\ref{fig:maps-physical}. FITS files with the maps are also available for
download, with names following the convention:
\texttt{KNNNN\_base.fits}. The file names always start with \texttt{K}, and
\texttt{NNNN} is the CALIFA ID, padded with zeros up to 4 digits. The suffix
\texttt{base} indicates the base used in the fit, which can be \texttt{gsd6e},
for base \textit{GMe}, or \texttt{zca6e}, for base \textit{CBe}.
For example, the galaxy shown in Fig.\ \ref{fig:maps-physical}, fitted using
\textit{GMe}, is stored as \texttt{K0140\_gsd6e.fits}. Tables with integrated
data fitted with both bases are also available for download, with the same
suffixes indicating which base.

There is also a form for selecting galaxies based on some of their
characteristics. These may be observational (Hubble type, SDSS magnitudes, etc.)
or physical quantities obtained on integrated spectra (stellar mass, mean
stellar age, $\tau_V$, etc.). Given that the dataset is small (445 galaxies
$\times 2$ bases), a simple search form suffices.

The observed data used as input for this work are available at the CALIFA web
site (\url{http://califa.caha.es/DR3/}).

\subsection{Description of the physical maps}
\label{sec:physical-maps}
\begin{figure*}
\centering
\includegraphics{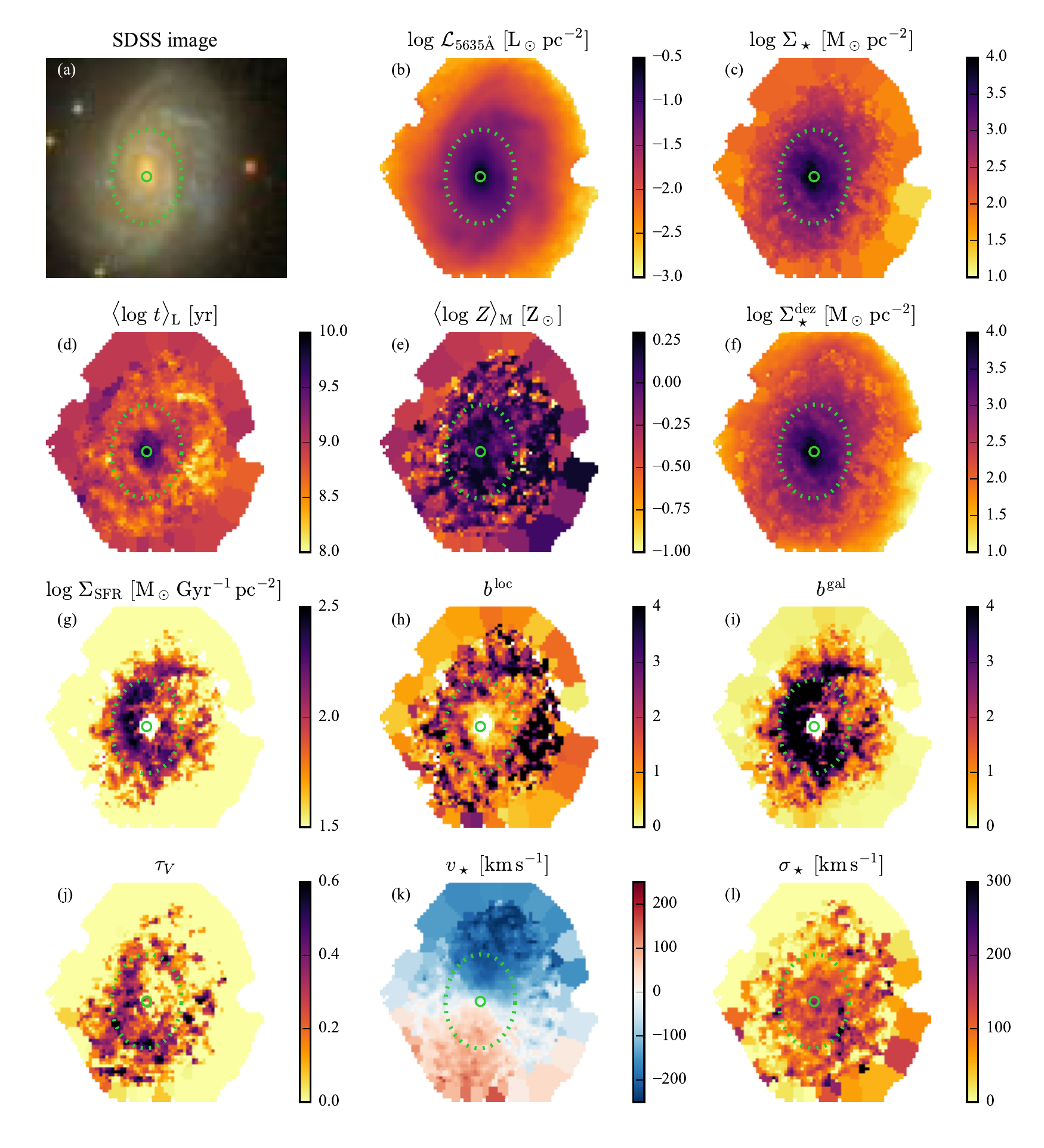}
\caption{Maps with physical properties of CALIFA 0140. A dotted
ellipse marks the $1\,\mathrm{HLR}$ area, and a small circle marks the galaxy
center. See Table \ref{tab:maps} for a brief description of each map.}
\label{fig:maps-physical}
\end{figure*}

To facilitate the description and illustrate the information content of the
catalogue, Fig.\ \ref{fig:maps-physical} shows the maps for a sample galaxy,
CALIFA 0140 (NGC 1667). The SDSS image\footnote{SDSS DR13 image obtained from
SkyServer.} of this galaxy is shown in panel (a). This is a low inclination Sbc
galaxy, with a small bar. It has a half light radius (HLR\footnote{The half
light radius (HLR) is straightforwardly computed by finding the distance where
the cumulative sum of the luminosity reaches half of its total value. This is
done along the major axis of ellipses with ellipticity and position angle
defined by the moments of the image \citep{Stoughton.etal.2002a}. Note that in
this work we calculate HLR using the $\mathcal{L}_{5635\text{\AA}}$ map, which
may or may not have masked regions.
Also, in some cases the FoV does not cover the whole galaxy. The HLR is a
convenient spatial metric, particularly useful when comparing results obtained
for different galaxies.}, computed in the $5635 \pm 45\,\text{\AA}$
normalization band) of $15.1^{\prime\prime}$, with an ellipticity of $\epsilon =
0.26$ and position angle of $177^\circ$.

In what follows we describe each of the maps available, as well as ways of
processing them to improve the results or express them in other forms to
highlight certain aspects of the data.

\subsubsection{Luminosity and stellar mass surface densities}
\label{sec:lum-mass}
The luminosity surface density ($\mathcal{L}_{5635\text{\AA}}$, Fig.\
\ref{fig:maps-physical}b) associated to each spaxel is calculated directly from
the observed spectra using the median flux in the spectral window of $5635 \pm
45\,\text{\AA}$. The map has units of
$\mathrm{L}_\odot\,\text{\AA}^{-1}\,\mathrm{pc}^{-2}$. The bulge and spiral arms
are clearly visible in the image.

The stellar mass surface density ($\Sigma_\star$, Fig.\
\ref{fig:maps-physical}c) is calculated using the masses derived from
\starlight, and are provided in units of $\mathrm{M}_\odot\,\mathrm{pc}^{-2}$.
This mass has been corrected for mass that returned to the interstellar medium
during stellar evolution, so it reflects the mass currently in stars. The
uncorrected mass, i.e., the total mass ever turned into stars,
$\Sigma^\prime_\star$, is also included in the database.

The morphological structures are still visible in the $\Sigma_\star$ map.
Nonetheless, an artifact is evident in the $\Sigma_\star$ map: the outer parts
of the galaxy are broken in flat patches. This happens because we combine the
lower S/N spaxels that are spatially close into Voronoi zones before running
\starlight. This behavior occurs in all maps that are products of the synthesis.
Only $\mathcal{L}_{5635\text{\AA}}$, which comes directly from the spectra, is
presented as a full resolution image.

\subsubsection{Dezonification}
\label{sec:dezonification}
Before proceeding with the description of the catalogue maps, it is useful to
present a simple data processing to smooth over the spatial effects of the
Voronoi zones used in the spectral fitting, as those seen in the image of
$\Sigma_\star$ (Fig.\ \ref{fig:maps-physical}c). This procedure was first
introduced by \citet{CidFernandes.etal.2013a} to smooth extensive properties,
and is generalized below.

The missing information on variations of  $\Sigma_\star$ within the spaxels in a
Voronoi zone can be approximately recovered using the
$\mathcal{L}_{5635\text{\AA}}$ image and some mass-to-luminosity relation. We
call this procedure ``dezonification''. In this paper, we calculate the mean
luminosity surface density in each zone, and divide each pixel of the
$\mathcal{L}_{5635\text{\AA}}$ map by this zone average. This yields a map of
weights, $w_{xy}$. More generally, one may compute weights raising the
luminosity to some {\em ad hoc} power $\alpha$. That is, for a given zone,
\begin{equation}
\label{eq:dezonification}
w_{xy} = \frac{\mathcal{L}_{xy}^\alpha} { N^{-1}
\sum\limits_{xy\,\in\,z} \mathcal{L}_{xy}^\alpha},
\end{equation}

\noindent where $x$ and $y$ are the coordinates of the spaxels inside the zone,
$N$ is the number of spaxels in the zone, and $\mathcal{L}_{5635\text{\AA}}$
is abbreviated to $\mathcal{L}$ for clarity. The single $\Sigma_\star$ value for
this zone can then be converted to an image through
\begin{equation}
\Sigma_{\star,xy}^\mathrm{dez} = w_{xy}\,\Sigma_{\star}. 
\end{equation}

\begin{figure}
\centering
\includegraphics{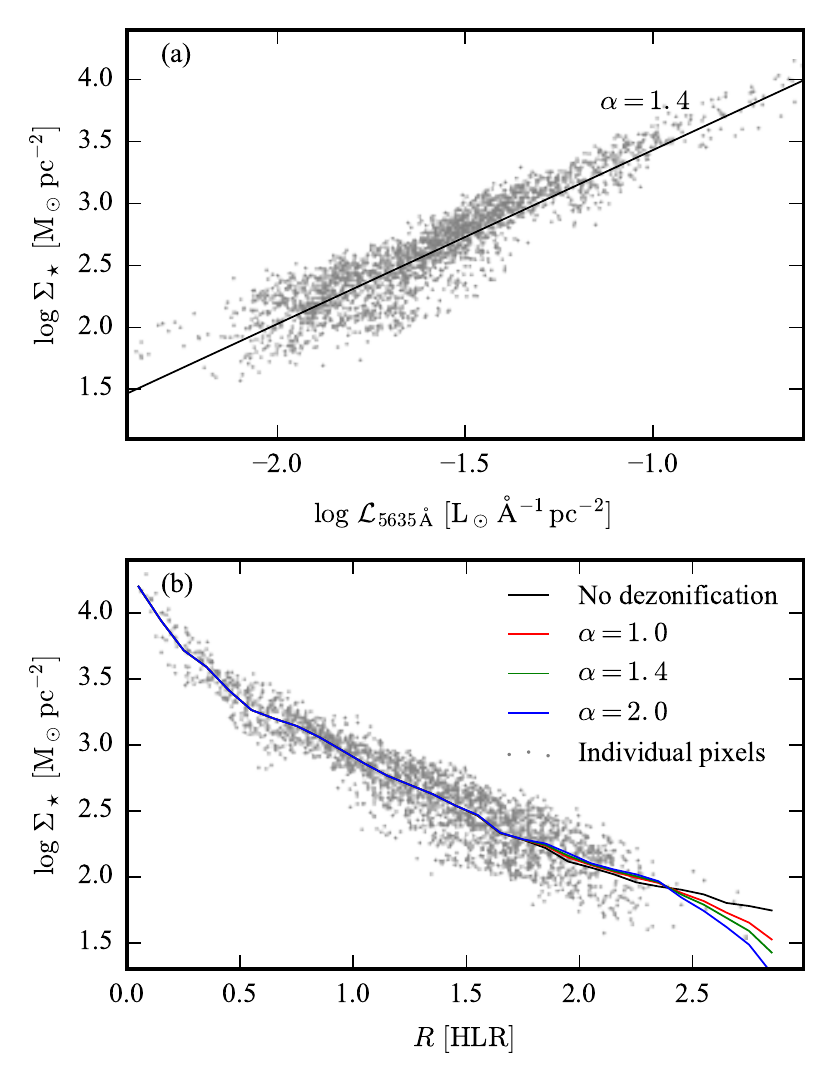}
\caption{Dezonification (see Section \ref{sec:lum-mass}) of the mass
surface density for CALIFA 0140. {\em Panel (a)} Mass--luminosity
plot to derive the $\alpha$ parameter used in Equation \ref{eq:dezonification}.
The black line is the linear fit to the data (gray dots). {\em Panel (b)} Radial
profile of the mass surface density. The dots are the mean mass surface density
in each zone. The black line is the mean radial profile without dezonification,
while the colored lines are the profiles with dezonification for different $\alpha$
values. The value used in this work is $\alpha = 1.4$ (green).}
\label{fig:dezonification}
\end{figure}

By design, the total mass of the zones does not change, we only redistribute the
mass among their pixels. Fig.\ \ref{fig:maps-physical}f shows an example of
dezonification using $\alpha = 1.4$. This value was found by fitting
$\Sigma_\star$ as a function of $\mathcal{L}_{5635\text{\AA}}$ for our example
galaxy, as seen in Fig.\ \ref{fig:dezonification}a. In Fig.\
\ref{fig:dezonification}b, we show radial profiles of $\Sigma_\star$, the
original (dots and black line) and dezonified versions with various values of
$\alpha$.
Inwards of $\sim 1.7\,\mathrm{HLR}$, the curves are identical, because zones
there actually correspond to single spaxels ($N = 1$). At larger radii the
$\alpha = 1.4$ line (in green) seems to better represent the cloud of dots for
this galaxy, while without dezonification (black line) the profile becomes
artificially flatter.
Note that the value for $\alpha$ was calculated for this galaxy only. Other
values may be needed for other galaxies, and can be straighforwardly obtained
with the data provided in the catalogue.
The same overall procedure can be applied to any property that is extensive,
that is, scales with the size of the system (for instance, $\Sigma^\prime_\star$
and $\Sigma_\mathrm{SFR}$). Intensive properties, like age and metallicity, are
inherently ``undezonifiable''.

\subsubsection{Mean stellar age and metallicity}
As amply documented in the \starlight-related literature (and indeed for any
stellar population synthesis method), the most robust way to summarise the
detailed star formation histories retrieved by the spectral fitting is through
the first moments of the age and metallicity distributions.  We present the mean
stellar age (in years) and metallicity (in solar units), both weighted by
luminosity and mass, defined as follows:
\begin{equation}
\label{eq:at_flux}
\langle \log\,t \rangle_\mathrm{L} = \sum_j x_j\,\log\,t_j
\end{equation}
\begin{equation}
\label{eq:at_mass}
\langle \log\,t \rangle_\mathrm{M} = \sum_j \mu_j\,\log\,t_j
\end{equation}
\begin{equation}
\label{eq:aZ_flux}
\langle \log\,Z \rangle_\mathrm{L} = \sum_j x_j\,\log\,Z_j
\end{equation}
\begin{equation}
\label{eq:aZ_mass}
\langle \log\,Z \rangle_\mathrm{M} = \sum_j \mu_j\,\log\,Z_j
\end{equation}

The luminosity weighted mean stellar age  (Equation \ref{eq:at_flux}) and mass
weighted mean stellar metallicity  (Equation \ref{eq:aZ_mass}) are shown in
panels (d) and (e) of Fig.\ \ref{fig:maps-physical}.
Most of the works on stellar population use these two definitions. Stellar
metallicity is directly related to the mass, which justifies this choice.
On the other hand, the mean age spans a larger dynamical range weighting by
light, which helps better mapping younger populations. A spiral pattern is
present in both maps, following the spiral seen in luminosity.
Negative gradients are present in both age and metallicity. These become more
evident with a little processing of the data, as seen in Section
\ref{sec:example-single}.

These are all intensive quantities, independent of the size of the system. Thus,
unlike $\Sigma_\star$ and $\Sigma_\mathrm{SFR}$, they are intrinsically not
dezonifiable.

\subsubsection{Star formation rate surface density}
Fig.\ \ref{fig:maps-physical}g shows the map of the mean recent star formation
rate surface density\footnote{Also known as star formation rate intensity in
some works.} ($\Sigma_\mathrm{SFR}$), obtained by adding the mass turned into
stars in the last $t_\mathrm{SF} = 32\,\mathrm{Myr}$ and dividing by this time
scale \citep{Asari.etal.2007a, CidFernandes.etal.2015a}. The maps are in units
of $\mathrm{M}_\odot\,\mathrm{Gyr}^{-1}\,\mathrm{pc}^{-2}$.
Close to the spiral arms we see higher values of $\Sigma_\mathrm{SFR}$.

Unlike for the mass and luminosity densities, this map has some empty patches.
To remove possible spurious young populations (intrinsic to the way \starlight
works), we mask the zones where the fraction of luminosity in populations
younger than $32\,\mathrm{Myr}$, that is, the $x_\mathrm{Y}$ map, is smaller
than $3\%$ (see \citealt{GonzalezDelgado.etal.2016a} for details).
Dezonification of this map is also possible, but there is a caveat. The
luminosity map represents the median luminosity in the $5635 \pm 45\,\text{\AA}$
band, a wavelength which is not usually dominated by the $t < t_{\mathrm{SF}}$
populations which go into the computation of $\Sigma_\mathrm{SFR}$. Users may
thus prefer to use the image at a shorter wavelength (extractable from the raw
DR3 data) to obtain more suitable dezonification weights.

It may be useful to consider $\Sigma_\mathrm{SFR}$ in relation to some fiducial
value instead of absolute units. \citet{CidFernandes.etal.2013a} uses Scalo's
birthrate parameter $b$ as an example, which measures the SFR in recent past ($t
< t_\mathrm{SF}$) with respect to the average over the entire lifetime of the
system. The latter is actually $\Sigma_\star$ divided by the age of the oldest
population in the base, $t_\infty = 14.1\, \mathrm{Gyr}$.
The birthrate parameter may be straightforwardly computed from the
$\Sigma^\prime_\star$ and $\Sigma_\mathrm{SFR}$ maps distributed in our
database:

\begin{equation}
b = \frac{\langle \Sigma_\mathrm{SFR} \rangle (t_\mathrm{SF})}
{\langle \Sigma_\mathrm{SFR} \rangle (t_\infty) } =
\Sigma_\mathrm{SFR} \frac{t_\infty} {\Sigma^\prime_\star}.
\end{equation}

The reference SFR value does not need to be local. For example, we define two
$b$ parameters, one where the reference value is the mean SFR surface density of
the same spaxel ($b^\mathrm{loc}$), and another where the reference value is the
mean SFR surface density of the whole galaxy ($b^\mathrm{gal}$). These maps are
shown in Figs.\ \ref{fig:maps-physical}h and \ref{fig:maps-physical}i.
$b^\mathrm{loc}$ shows regions where the star formation is currently stronger
locally, independent of the total mass formed, while $b^\mathrm{gal}$ shows
where in the galaxy the current SFR is stronger. Note that the $b^\mathrm{gal}$
map is equivalent to that in Fig.\ \ref{fig:maps-physical}d, but in different
units and in a linear scale. These maps are shown here as examples of simple
operations over the data distributed in our catalogue to highlight one or
another aspect of interest.

\subsubsection{Attenuation coefficient}
\starlight models the dust attenuation with a foreground screen factor, that is
\begin{equation}
\label{eq:attenuation}
F^\mathrm{obs}_\lambda = F^\mathrm{int}_\lambda
\exp \left( -\tau_V\,q_\lambda \right),
\end{equation}

\noindent where $F^\mathrm{int}_\lambda$ is intrinsic flux behind the dust
screen, and $F^\mathrm{obs}_\lambda$ is the resulting observed flux. Fig.\
\ref{fig:maps-physical}j shows the map of the V-band optical depth $\tau_V$
obtained for CALIFA 0140. The values range from $0$ to $0.6$, with an overall
average of $0.1$. The morphology of the $\tau_V$ maps roughly follows that of
the recent SFR (panels d and f), which fits with the notion that star-forming
regions are dustier than the general interstellar medium.

As in the case of mean ages and metallicities, the $\tau_V$ maps are not
dezonifiable. One could, in principle, modify our dezonification scheme by
defining weights in terms of colors instead of fluxes, and attribute larger
$\tau_V$ values to the redder spaxels within a zone according to an empirically
calibrated relation analogous to that in Fig.\ \ref{fig:dezonification}a. Known
degeneracies with other stellar population properties (mainly age) complicate
this extension of the dezonification method, but it might still be worth
exploring this possibility.

\subsubsection{Stellar kinematics}
Stellar velocity ($v_\star$) and velocity dispersion ($\sigma_\star$) maps are
shown in Figs.\ \ref{fig:maps-physical}k and l, respectively.
Again, dezonification is not possible with these images.

The rotation pattern is clearly seen in the $v_\star$ map, with projected speeds
reaching $\sim 200\,\mathrm{km\,s}^{-1}$. The $\sigma_\star$ map shows a decline
towards the outside. There are, however, several low $\sigma_\star$ patches, as
well as some high $\sigma_\star$ spikes, none of which seem real.

These artifacts are a consequence of the low resolution
($\mathrm{FWHM}=6\,\text{\AA}$) of the spectra, as well as of intrinsic
difficulties faced by \starlight in fitting $\sigma_\star$ under some
circumstances (like noisy spectra or dominant young populations). We thus
recommend caution in the use of the $\sigma_\star$ maps provided in the
catalogue. A much more precise kinematical analysis is possible with the V1200
grism ($\mathrm{FWHM}=2.7\,\text{\AA}$) datacubes, not analysed here because
their limited wavelength coverage makes them less informative for a stellar
population analysis. \citet{FalconBarroso.etal.2017a} have recently
published\footnote{
Stellar kinematics maps from \citet{FalconBarroso.etal.2017a} are available at
\url{http://www.caha.es/CALIFA/public_html/?q=content/science-dataproducts}.} a
detailed kinematical analysis of 300 CALIFA galaxies using this higher
resolution setup.

\subsection{Quality-control and uncertainties}
\label{sec:quality-control}
Despite the homogeneity of CALIFA data, there are relevant variations in data
quality both within and between galaxies. Moreover, the \starlight fits
themselves can be bad even in cases the data are good, e.g., because of unmasked
strong emission lines. It is thus important to keep track of these caveats in
order to evaluate to which extent they affect the results of any one particular
analysis.

Our catalogue offers a suite of indices to carry out this quality control. The
indices are listed in rows 12--18 of Table \ref{tab:maps}. Quality control maps
for CALIFA 0140 are shown as an example in Fig.~\ref{fig:maps_diag} and
discussed in sections \ref{sec:data-quality} and \ref{sec:fit-quality} below.
A discussion of the uncertainties in the derived properties is also presented
(section \ref{sec:uncertainties}).

\subsubsection{Data quality maps}
\label{sec:data-quality}
\begin{figure*}
\centering
\includegraphics{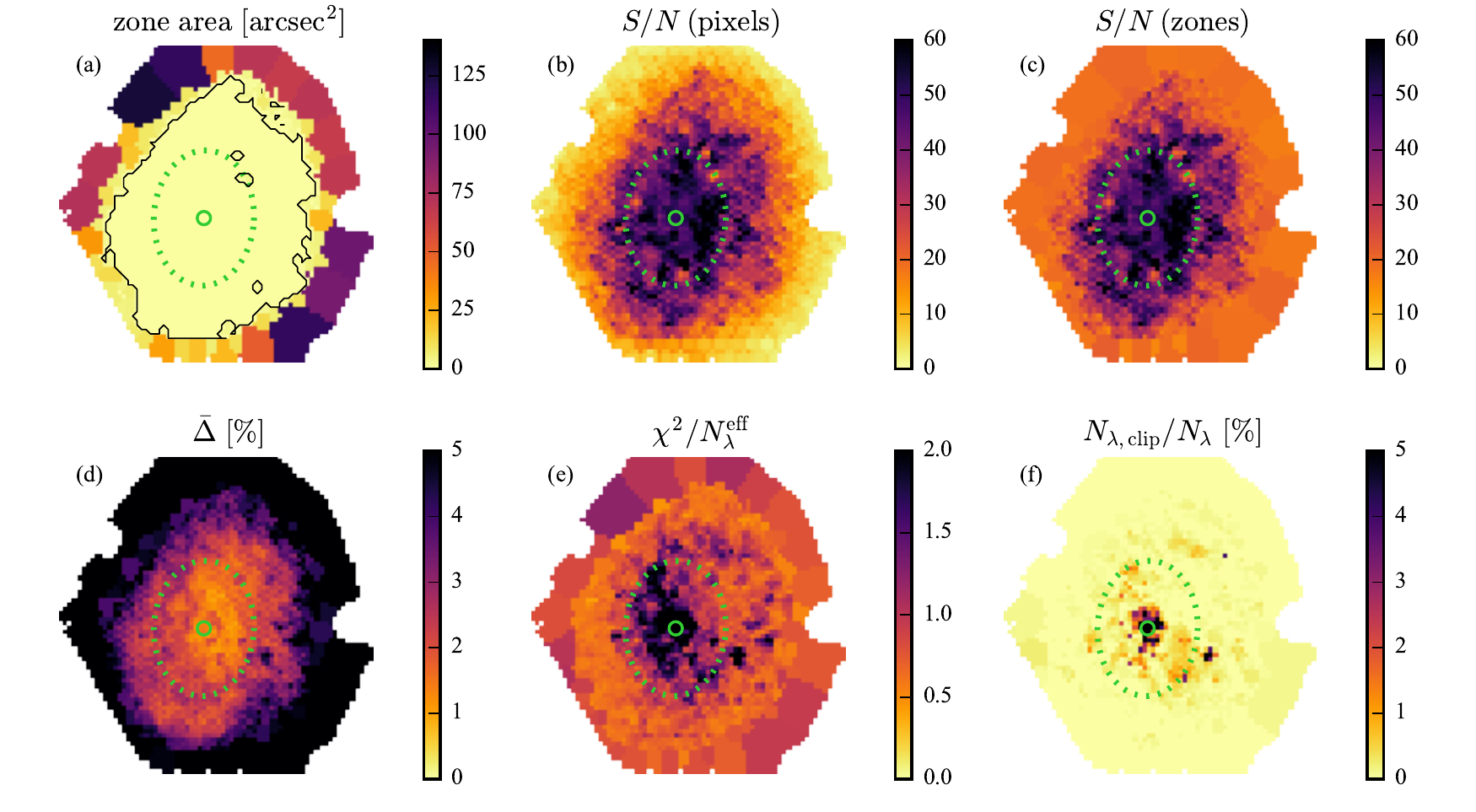}
\caption{Maps with fit quality indicators of NGC 1667 (CALIFA 0140). A dotted
ellipse marks the $1\,\mathrm{HLR}$ area, and a small circle marks the galaxy
center. See Table \ref{tab:maps} for a brief description of each map.}
\label{fig:maps_diag}
\end{figure*}

Before running the spectra through \starlight, some pre-processing is necessary.
The most basic is flagging spaxels that we do not want to apply the synthesis
to. That includes spaxels with artifacts, foreground objects, and very low S/N
spaxels ($< 3$). These become masked regions in the maps, where there is no
data, appearing white in Figs.\ \ref{fig:maps-physical} and \ref{fig:maps_diag}.
The mask is stored in the Bad pixel map.

As discussed in Section \ref{sec:preprocessing}, and detailed in section 3.2 of
\citet{CidFernandes.etal.2013a}, the spaxels with S/N lower than $20$ are then
binned into Voronoi zones. The zones map assigns a zone number to each pixel, so
that if two or more pixels are contained in a given zone, then they share the
same number.

Fig.\ \ref{fig:maps_diag}a shows the area of the Voronoi zones for our example
galaxy. As expected, the outer parts of the galaxy have larger zones. The black
contour indicates the region where the zones are composed of a single pixel,
reaching approximately $2\,HLR$ of the galaxy in this case. This map may be
used, for example, to calculate the total mass in each zone. It is also used in
the dezonification procedure explained in Section \ref{sec:dezonification}.

Fig.\ \ref{fig:maps_diag}b shows the pixel-by-pixel S/N ratio in the $5635 \pm
45\,\text{\AA}$ band. The outer parts of the galaxy have S/N well below $10$.
With the Voronoi binning all the spectra have $\mathrm{S/N} > 20$, as can be
seen in Fig.\ \ref{fig:maps_diag}c.

\subsubsection{Fit quality maps}
\label{sec:fit-quality}
Even if the spectra were classified as good by the previous criteria, we must
assert that the fit was successful. Fig.\ \ref{fig:maps_diag}d shows the map of
the fit quality indicator $\overline{\Delta}$, defined as
\begin{equation}
\overline{\Delta} = \frac{1}{N_\lambda^\mathrm{eff}} \sum_\lambda
\frac{\left| O_\lambda - M_\lambda \right|} {M_\lambda},
\end{equation}

\noindent where $N_\lambda^\mathrm{eff}$ is the number of wavelengths
effectively used in the fit, and $O_\lambda$ and $M_\lambda$ are the observed
and model spectra, respectively. The figure of merit actually used by \starlight
is $\chi^2$, which is shown in Fig.\ \ref{fig:maps_diag}e divided by
$N_\lambda^\mathrm{eff}$.

Usually, one would resort to $\chi^2$ to check the quality of the fit. However,
$\chi^2$ is closely tied to the uncertainty of the spectra ($\epsilon_\lambda$),
which is sometimes difficult to estimate precisely. Due to the clipping
algorithm, discussed below, the number of measurements effectively used in the
fit ($N_\lambda^\mathrm{eff}$) varies between the spaxels. Therefore we use
$\chi^2/N_\lambda^\mathrm{eff}$ to be able to compare the fit statistic between
them. Inspection of Fig.\ \ref{fig:maps_diag}e may lead to the wrong conclusion
that the spectral fits are worse in the nuclear regions than outside, but the
larger $\chi^2/N_\lambda^\mathrm{eff}$ values in those regions is because of
smaller errors (larger S/N, as seen in panel c), not worse fits. The mean
absolute deviation $\overline{\Delta}$ does not depend explicitly on the
uncertainties, so its map in Fig.\ \ref{fig:maps_diag}a conveys a more
appropriate measure of the fit quality.

Another useful quantity to keep track of is the amount of clipping done by
\starlight during the fit. If some wavelengths have fluxes that are outliers,
over $\pm 4\,\epsilon_\lambda$ off the model during the fit, they are flagged as
clipped, and not used in the fit anymore. This is enforced to keep artifacts and
non-masked emission lines from messing up the fit. The fraction of clipped
wavelengths, given by $N_{\lambda,\mathrm{clip}} / N_\lambda$, is shown in Fig.\
\ref{fig:maps_diag}f. For CALIFA 0140, the median fraction of clipping is
$3.5\%$, with very few going over $5\%$. Nevertheless, given that these are
usually only a handful among the thousands of pixels, they are usually
irrelevant for statistics that take large slices of the maps.

\subsubsection{Uncertainties}
\label{sec:uncertainties}

\starlight does not provide error estimates in its output. These uncertainties
are usually explored by perturbing the data according to some prescription
related to observational and calibration errors, and performing the spectral
synthesis several times \citep[e.g.,][]{CidFernandes.etal.2005a}.

It is not practical, in our case, to run the synthesis code dozens of times for
all spectra from CALIFA. We can, however make use of the simulations by
\citet{CidFernandes.etal.2014a} to estimate the uncertainties associated to the
physical properties obtained by \starlight. The simulations were performed using
data from a single galaxy, CALIFA 0277. The spectra were perturbed in different
ways to estimate the effects of noise, calibration issues and multiplicity of
solutions.

\begin{table}
	\caption{Estimated uncertainties of physical quantities, obtained by
	simulations \citep{CidFernandes.etal.2014a}.}
	\label{tab:uncertainties}
	\centering
	\begin{tabular}{c c c}
		\hline\hline
		        & \multicolumn{2}{c}{Uncertainty} \\
		Quantity & $20 < \mathrm{S/N} < 30$ & $40 < \mathrm{S/N} < 50$ \\
		\hline
		$\log\,M_\star$ & $0.09$ & $0.04$ \\
		$\langle \log\,t \rangle_\mathrm{L}$ & $0.08$ & $0.05$ \\
		$\langle \log\,t \rangle_\mathrm{M}$ & $0.15$ & $0.07$ \\
		$\langle \log\,Z \rangle_\mathrm{L}$ & $0.08$ & $0.05$ \\
		$\langle \log\,Z \rangle_\mathrm{M}$ & $0.12$ & $0.07$ \\
		$\log\,\mathrm{SFR}$                 & $0.16$ & $0.14$ \\
		$x_\mathrm{Y}$                       & $0.03$ & $0.01$ \\
		$\tau_V$                             & $0.05$ & $0.02$ \\
		$v_\star\,[\mathrm{km}\,\mathrm{s}^{-1}]$      & $19$ & $9$ \\
		$\sigma_\star\,[\mathrm{km}\,\mathrm{s}^{-1}]$ & $22$ & $10$ \\
		\\
		\hline
	\end{tabular}
\end{table}

Table \ref{tab:uncertainties} lists estimated uncertainties associated to the
maps in our catalogue. These values are based on the OR1 simulations in
\citet{CidFernandes.etal.2014a}, in which Gaussian noise with amplitude given by
the error in fluxes was injected in the data. The uncertainties are obtained by
taking the standard deviation of the difference between the results from
perturbed and original spectra.
Except for $x_Y$, $v_\star$ and $\sigma_\star$, the uncertainties are
logarithmic, independent of scale. That means the uncertainties in $\log
\Sigma_\star$ and in $\log \Sigma^\prime_\star$ are the same as the
uncertainties for $\log M_\star$ given in the table.

Table \ref{tab:uncertainties} complements and expands upon a similar table
presented in \citet{CidFernandes.etal.2014a}. While in that paper the
uncertainties were evaluated considering all zones of the galaxy, here we break
them into two bins in S/N, one for zones where $20 < \mathrm{S/N} < 30$, and
another one for $40 < \mathrm{S/N} < 50$, which roughly translates to outer and
inner regions of galaxies, respectively (see Fig.~\ref{fig:maps_diag}c for an
example). We find that uncertainties decrease by typically a factor of two from
the high to the low  S/N bin. Uncertainties for individual zones may be obtained
by interpolating the values given in Table \ref{tab:uncertainties}, and using
the S/N diagnostic maps. Naturally, these uncertainty estimates must be
interpreted as approximate, as they are based on experiments with a single
galaxy, and do not take into consideration sources of error other than random
noise (like the continuum calibration, also explored by
\citealt{CidFernandes.etal.2014a}).

An independent and more empirical estimate of the uncertainties may be obtained
from the catalogue data itself, by comparing regions known to have similar
spectra, as those along fixed radial bins in elliptical galaxies, or in the
bulges of spirals. This recipe is explored in Section \ref{sec:radprof} for
CALIFA 0140.

\subsection{Integrated spectral analysis}
We now turn to the properties derived from the integrated spectra in our
catalogue (i.e.\ simulating a single-beam observation for each galaxy).
In the following, we compare physical properties obtained directly from the
integrated spectra to the same properties obtained by integrating the spatially
resolved results. Since there is only one integrated spectrum per object, we
must thus set aside the analysis of one single sample galaxy and consider all
galaxies in our main sample.

\begin{figure*}
\centering
\includegraphics{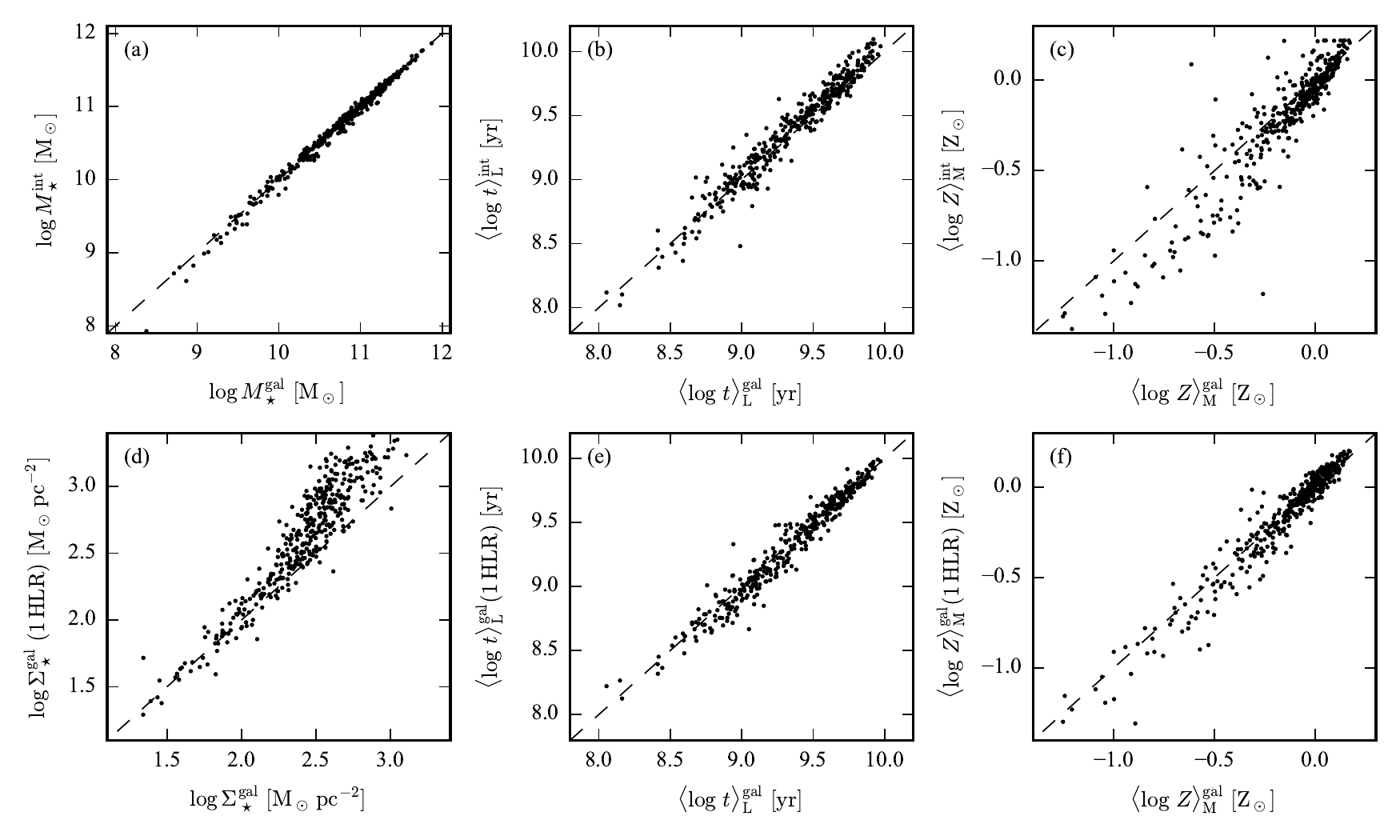}
\caption{Properties from integrated spectra (superscript ``int'') {\em versus}
spatially resolved properties (superscript ``gal'') for all galaxies in the
sample. {\em Panel (a)} Mass obtained from the integrated spectra against the
total mass in the integral of mass surface density map. {\em Panel (b)}
Luminosity-weighted stellar age from the integrated spectra against the mean of
stellar ages of all pixels. {\em Panel (c)} The same as panel (b), but for
mass-weighted stellar metallicity. {\em Panel (d)} Mean stellar mass surface
density of the whole galaxies against the mean in a ring of $1\,\mathrm{HLR}$
around the galaxy nucleus. {\em Panel (e)} The same as panel (d), but for
luminosity-weighted stellar age. {\em Panel (f)} The same as panel (d), but for
mass-weighted stellar Metallicity. The dashed lines show the $y=x$ diagonal.}
\label{fig:integrated}
\end{figure*}

Fig.~\ref{fig:integrated}a shows that the total stellar mass $M_\star$ obtained
from the integrated spectra and integrated spatially resolved masses agree
remarkably well.

For intensive properties, the way to reach a global value will vary according to
the nature of the property. Average ages and metallicities must be weighted by
luminosity or mass. For instance, taking into consideration Equation
\ref{eq:at_flux}, we may calculate the average stellar age of the whole galaxy
(in fact, for any set of pixels), weighted by luminosity, using
\begin{equation}
\label{eq:weighted-mean}
\langle \log\,t \rangle^\mathrm{gal}_\mathrm{L} = \frac{\sum_{x,y}
\mathcal{L}_{xy}\,\langle \log\,t \rangle_{\mathrm{L},xy}} {\sum_{x,y}
\mathcal{L}_{xy}},
\end{equation}

\noindent where $\mathcal{L}_{xy}$ and $\langle \log\,t \rangle_{\mathrm{L},xy}$
are the maps for $\mathcal{L}_{5635\text{\AA}}$ and $\langle \log\,t
\rangle_\mathrm{L}$, respectively. The weight for mass weighted quantities is
$\Sigma_{\star,xy}$. The same equation applies to stellar metallicity. Figs.\
\ref{fig:integrated}b and \ref{fig:integrated}c shows the comparison between the
integrated and mean of the maps for $\langle \log\,t \rangle_\mathrm{L}$ and
$\langle \log\ Z \rangle_\mathrm{M}$, respectively. There is a fairly good
agreement of both measurements for mean stellar age, with a bit more scatter
than the plot for stellar mass. The results for metallicity are similar, but
with more scatter.

In the lower panels of Fig.\ \ref{fig:integrated} we present a comparison
between the mean values over the whole galaxy to the mean over a ring at $1 \pm
0.1\,\mathrm{HLR}$ from the nucleus. The plots show that, for stellar ages and
metallicities, the value at $1\,\mathrm{HLR}$ is a good proxy for the mean of
the whole galaxy. The same is not strictly valid for $\Sigma_\star$ (Fig.\
\ref{fig:integrated}). This happens because the mean stellar mass surface
density of a galaxy is not a well defined value. Given that $\Sigma_\star$
decreases with the radius, depending on which radius is chosen as the
``boundary'' of the galaxy, the surface density can become arbitrarily small.
The galaxies of the sample were observed typically up to $\sim
2.5\,\mathrm{HLR}$. Calculating the average over $2\,\mathrm{HLR}$, for example,
moves the points closer to the diagonal. These results confirm what was obtained
by \citet{GonzalezDelgado.etal.2014b}, who used the first 107 galaxies observed
by CALIFA to perform this same kind of integrated versus resolved analysis.

\subsection{Comparison of results with different bases}
\label{sec:GMxCB}

\begin{figure*}
\centering
\includegraphics{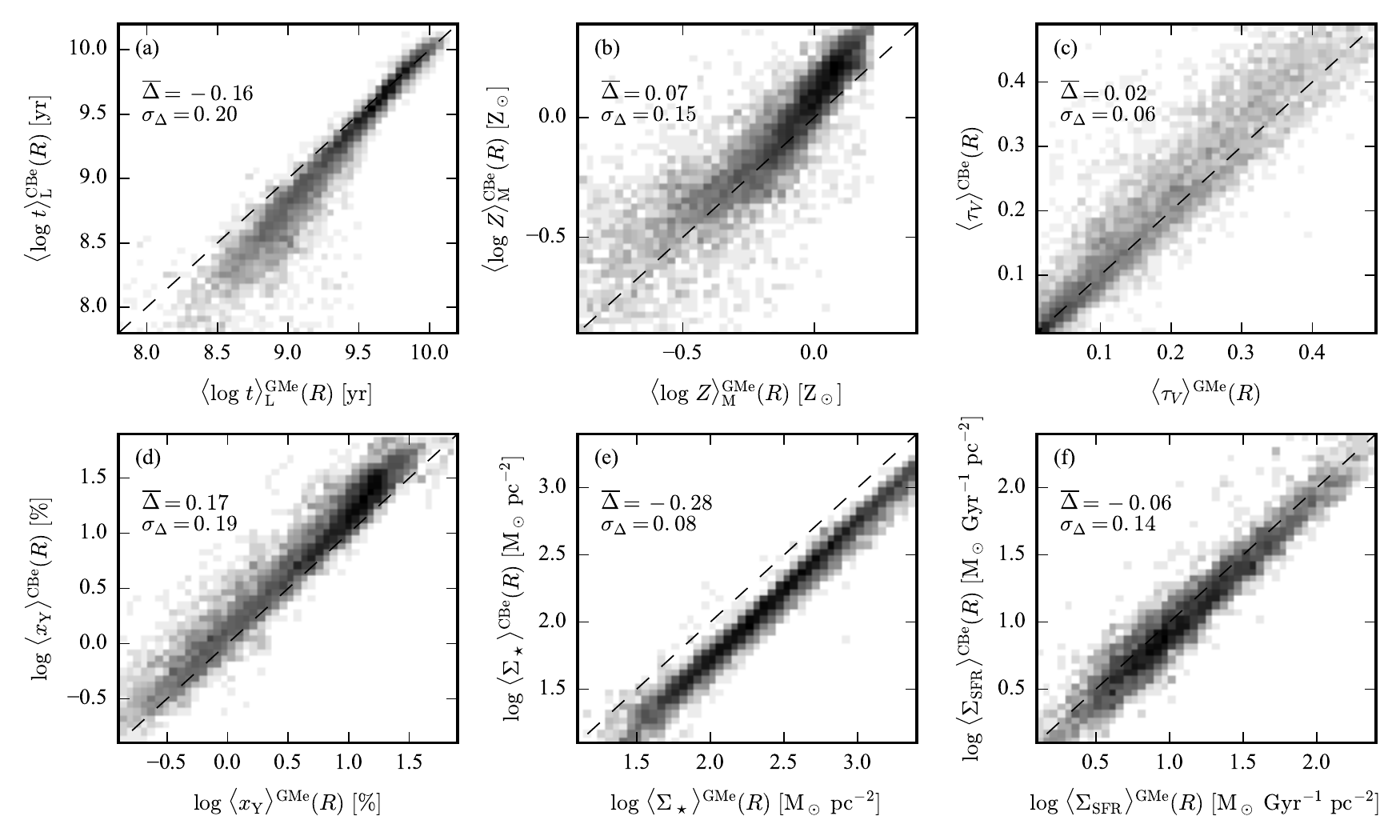}
\caption{Comparison between the synthesis results using bases
\textit{GMe} and \textit{CBe}. The panels show 2D histograms of values
averaged in radial bins of all 395 maps of the main sample. The horizontal and
vertical axes are values obtained using \textit{GMe} and \textit{CBe},
respectively. In each panel we also show the mean and standard deviation of the
difference between \textit{CBe} and \textit{GMe}.
{\em  (a)} Luminosity weighted mean stellar age.
{\em  (b)} Mass weighted mean stellar metallicity.
{\em  (c)} Dust attenuation.
{\em  (e)} Fraction of light in populations younger than $32\,\mathrm{Myr}$.
{\em  (e)} Stellar mass surface density.
{\em  (f)} Star formation rate surface density.
}
\label{fig:GMxCB}
\end{figure*}

All examples presented in this paper were obtained with base \textit{GMe},
described in Section \ref{sec:spec-synthesis}. As mentioned there, our catalogue
also offers results derived with \textit{CBe}. Previous papers in this series
combining CALIFA data with a \starlight analysis have compared results obtained
with these two bases, drawn from independent sets of evolutionary synthesis
models
(\citealt{CidFernandes.etal.2014a,GonzalezDelgado.etal.2015b,GonzalezDelgado.etal.2016a}).
In the interest of completeness, and also to update these previous comparisons
to our final sample, Fig.\ \ref{fig:GMxCB} compares properties derived with
these two bases, for all 395 galaxies of the main sample.
These properties were
averaged in radial bins up to $2.5\,\mathrm{HLR}$, in steps of
$0.1\,\mathrm{HLR}$. See Section \ref{sec:radprof} for more details on radial
profiles. The averaging strategy for intensive quantities is the same adopted in
the previous section, using Equation \ref{eq:weighted-mean}.

Panels a--d of Fig.\ \ref{fig:GMxCB} compare the intensive properties $\langle
\log\,t \rangle_\mathrm{L}$ (panel a), $\langle \log\,Z \rangle_\mathrm{M}$ (b),
$\tau_V$ (c), and the fraction of light in stellar populations younger than
$32\,\mathrm{Myr}$ ($x_\mathrm{Y}$, d) obtained with bases \textit{GMe}
(horizontal axis) and \textit{CBe} (vertical axis).
The other panels compare the results for $\Sigma_\star$ (panel e), and
$\Sigma_\mathrm{SFR}$ (f), both of which are extensive and IMF-sensitive
quantities. In all panels the mean ($\overline{\Delta}$) and standard deviation
($\sigma_\Delta$) of the $\Delta \equiv y - x = $ \textit{CBe} $-$ \textit{GMe}
difference are indicated.

The results in Fig.\ \ref{fig:GMxCB} agree with our previous comparisons for
smaller samples. Mean stellar ages using \textit{CBe} are $0.16$ dex lower, on
average. As indicated by the tendency of the offset to increase towards smaller
values of $\langle \log\,t \rangle_\mathrm{L}$ in panel a, the difference is
mostly due to the fact that young stellar populations come out somewhat stronger
with \textit{CBe} than with \textit{GMe}, as confirmed in panel d. Metallicities
are somewhat larger when using \textit{CBe} because its base models reach higher
$Z$ than \textit{GMe}, while the dust attenuation is essentially the same
($\overline{\Delta}\tau_V = 0.02$).
The stellar mass surface density is lower by a factor of $\sim 2$ when using
\textit{CBe}, reflecting the different IMF used in each base. Differences in
$\Sigma_\mathrm{SFR}$ values, however, are not as large, with an offset of only
$0.06$ dex. As seen above in the $x_\mathrm{Y}$ comparison, the fraction of mass
in young stellar populations is higher with \textit{CBe}, which compensates for
the lower absolute mass due to differences between the Chabrier and Salpeter
IMFs.

Despite the non-negligible differences between bases \textit{CBe} and
\textit{GMe}, the general results are essentially the same. There are,
naturally, systematic offsets in the properties derived with these two sets of
models, but these are ultimately not relevant in a comparative analysis, which
is essentially what one aims when studying spatial variations of properties such
as mean stellar age,  metallicity, or attenuation caused by dust.

\section{Using the catalogue}
\label{sec:using-the-catalogue}
The data in our \pycasso database can be used in a variety of ways to address a
wide ranging spectrum of questions. This section gives concrete examples of the
kind of work that can be carried out with some simple processing of the maps
made publicly available, as described in Section \ref{sec:data-access}.

An underlying characteristic of all examples below is that they explore the
statistical power of the database, be it using all 316,960 zones in all 395
galaxies from the main sample, in subsamples focusing on certain types of
galaxies, or even in single galaxies, which become statistical samples
themselves with IFS data. This is not only a sensible approach, but a firmly
established working philosophy within the stellar population synthesis
community, which, acknowledging the several uncertainties involved in properties
retrieved from single spectra, has long vouched for a statistical approach
\citep{Panter.etal.2007a, CidFernandes.etal.2013a}.

Examples of how to use our database for single galaxies are explored  in Section
\ref{sec:example-single}, while Section \ref{sec:example-many} presents examples
of what can be done for samples of galaxies. As anticipated in the introduction
(and as already done in Fig.\ \ref{fig:integrated}), we take the opportunity to
update some of the results obtained in our own earlier work.

\subsection{Examples of usage for a single galaxy}
\label{sec:example-single}
Maps such as those in Fig.\ \ref{fig:maps-physical} are good for a global
appreciation of the results and locating particular spots of the galaxies, like
HII regions or spiral arms, but they are not convenient to visualise trends and
gradients. By reducing the dimensionality of the data we may have a better grasp
of the behaviour of the physical properties. With radial profiles, we take
advantage of both statistics and visualisation.

\subsubsection{Radial profiles}
\label{sec:radprof}
\begin{figure}
\centering
\includegraphics{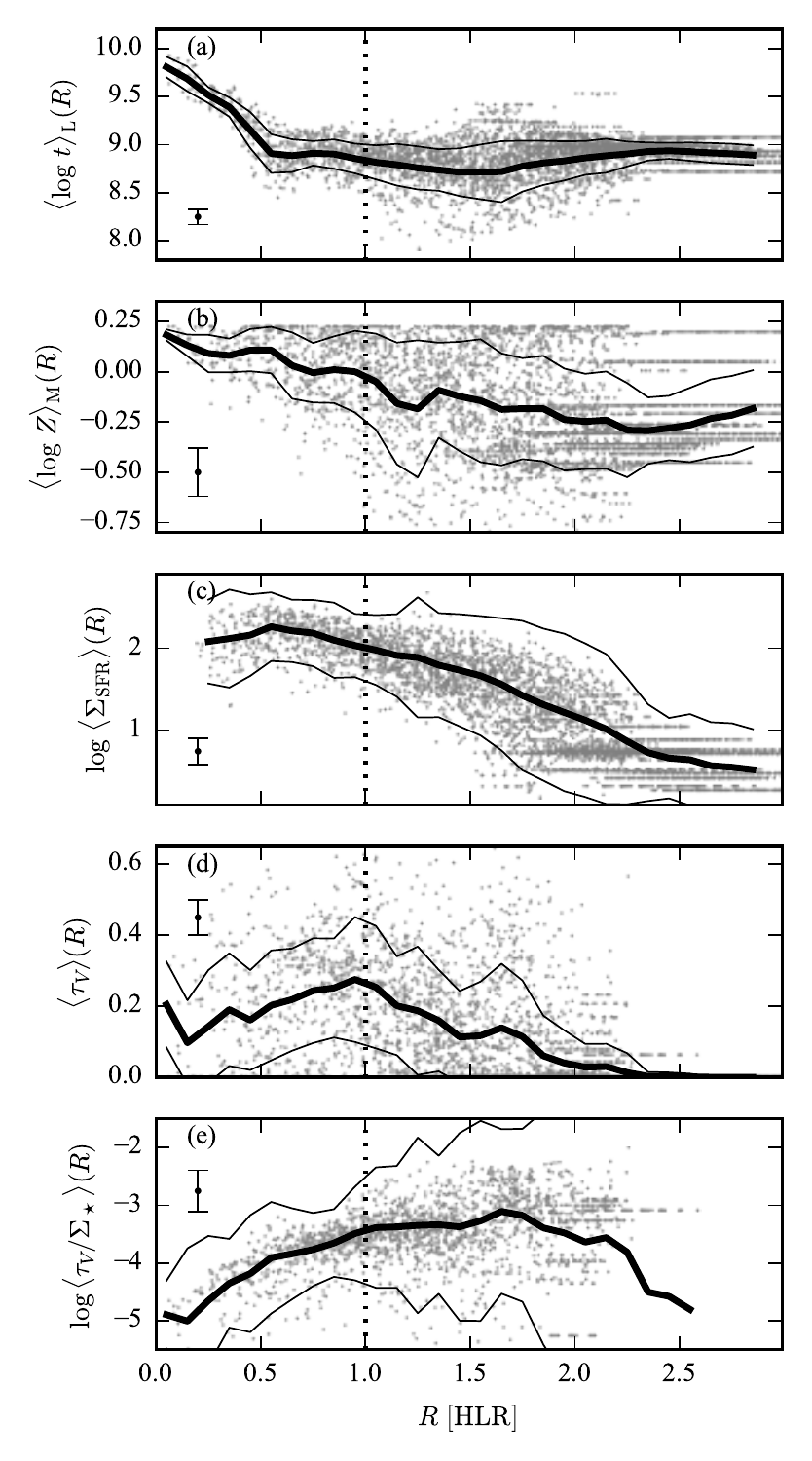}
\caption{Radial profiles of some physical quantities for CALIFA
0140. The gray dots are the values for individual pixels. The black lines are
the mean values in radial bins of width $0.1\,\mathrm{HLR}$, with the $1\sigma$
range shown by thin black lines. The vertical dotted line marks
$1\,\mathrm{HLR}$.
Typical uncertainties associated to these quantities (with the lowest S/N
from Table \ref{tab:uncertainties}) are shown as single error-bars at the left
side of the plots. See Table \ref{tab:maps} for a brief description of these
quantities.
From top to bottom: luminosity-weighted mean stellar age (a), in
$[\mathrm{yr}]$; mass-weighted mean stellar metallicity (b), in
$[\mathrm{Z}_\odot]$; star formation rate surface density (c), in
$[\mathrm{M}_\odot\,\mathrm{Gyr}^{-1}\,\mathrm{pc}^{-2}]$; dust attenuation
(d), dimensionless; dust--to--stars ratio (e), in arbitrary units.}
\label{fig:radprof}
\end{figure}

As an example, Fig.\ \ref{fig:radprof} shows radial profiles for some of the
physical properties for CALIFA 0140. Each plot shows the values of some property
for each zone as gray dots, while the average and $\pm$ one sigma dispersion in
radial bins are shown by the thick and thin solid lines. The radial bins are
elliptical annulli with thickness of $0.1\,\mathrm{HLR}$.
A vertical dashed line marks the HLR for this galaxy. We warn that this simple
statistics may not be the wiser way to treat the physical quantities in this
plot, but it suffices for our current illustrative purposes.

Panels a--d show the radial run of $\langle \log\,t \rangle_\mathrm{L}$,
$\langle \log\,Z \rangle_\mathrm{M}$, $\Sigma_{\mathrm{SFR}}$, and $\tau_V$,
respectively. The estimated uncertainties, described in Section
\ref{sec:uncertainties} are shown as single errorbars in each panel. There is
substantial scatter in the plots, but the mean radial profiles are well behaved.
The mean stellar age ($\langle \log t \rangle_\mathrm{L}$, Fig.\
\ref{fig:radprof}a shows a clear drop from the center down to $R \sim
0.7\mathrm{HLR}$, where it levels off at a value of about $0.5\,\mathrm{Gyr}$.
The mass weighted mean stellar metallicity $\langle \log\,Z \rangle_\mathrm{M}$
decreases more or less steadily by $\sim 0.4$ dex from the nucleus to the
outskirts of the galaxy (Fig.\ \ref{fig:radprof}b). The large scatter is not
surprising considering that metallicities are notoriously harder to estimate
than ages. The $\Sigma_\mathrm{SFR}$ profile (Fig.\ \ref{fig:radprof}c) peaks at
about $0.7\mathrm{HLR}$, coinciding with the inflection in the $\langle \log t
\rangle_\mathrm{L}$ profile, decreasing both towards the inner bulge region and
the outer disk, as indeed seen (but less clearly so) in the full
$\Sigma_\mathrm{SFR}$ maps in Fig.\ \ref{fig:maps-physical}g. The dust optical
depth\footnote{ Note that \starlight assumes a dust screen and finds the
\emph{effective} attenuation at each zone of a CALIFA galaxy. Thus one should be
careful when interpreting $\tau_V$ as the dust optical depth. In addition, the
arithmetic averaging of $\tau_V$ may be misleading: the radial profiles show the
typical value of $\tau_V$ at a given radius, but not the effective attenuation
one would obtain if analysing the coadded spectra in a ring.} shows weaker
trends, slowly decreasing towards $0$ outwards of $\sim 1\,\mathrm{HLR}$ (more
or less in tandem with the $\Sigma_\mathrm{SFR}$ profile), and slowly decreasing
to values around $0.2$ in the nucleus.

As anticipated in section \ref{sec:uncertainties}, these profiles may be used to
estimate the uncertainties adopted for the properties. For CALIFA 0140 the
region inside $0.7 \mathrm{HLR}$ is the bulge of the galaxy. Given that the
bulge is fairly symmetrical, the spectra in a radial bin will be roughly
equivalent to multiple observations of the same object. This means that the
dispersion of the radial profiles serve as an empirical estimate of the
uncertainty of these properties. In fact, as the error-bars in Fig.\
\ref{fig:radprof} illustrate, they are in good agreement with the estimated
uncertainties obtained by simulation \citep{CidFernandes.etal.2014a}, presented
in Table \ref{tab:uncertainties}.

The argument of symmetry does not apply to the outer parts of this particular
galaxy. The spiral pattern breaks the symmetry, which means the dispersion in
the profiles increases outside of $0.7\mathrm{HLR}$, is caused by the data.
Nevertheless it is still a good estimator of the uncertainty for, e.g.,
asserting that a gradient in a radial profile is significant.

Further analysis may be done by combining the values present in the catalogue
maps. For instance, in Fig.\ \ref{fig:radprof}e we show $\tau_V / \Sigma_\star$
to obtain a proxy for the dust-to-stars ratio (itself a proxy for the
gas-to-stars ratio given some hypothesis about dust/gas). As shown in
\citet{GonzalezDelgado.etal.2014b}, $\tau_V$ profiles generally tend to show
negative gradients, which one might read as a tendency for galaxies to be less
dusty towards their outer regions. Fig.\ \ref{fig:radprof}e conveys a different
impression, however. The rise of $\tau_V / \Sigma_\star$ with radius indicates
that here are more dust particles per star as one moves away from the nucleus, a
trend which may ultimately be reflecting the fact that galaxies are more gas
rich towards the outside.

\subsubsection{Bulge-disk decomposition in light and mass}
\label{sec:morph-fit}
Perhaps the most traditional type of surface photometry analysis for spirals is
bulge/disk decomposition \citep{Kormendy.1977a, Kent.1985a}.
The data distributed in our catalogue allow one to perform this kind of analysis
both in light ($\mathcal{L}_{5635\text{\AA}}$) and mass ($\Sigma_\star$) surface
densities.

A morphological  analysis of our $\mathcal{L}_{5635\text{\AA}}$ images (or those
produced at any other $\lambda$) is not particularly attractive {\em per se}, as
the image quality of CALIFA is inferior that of publicly available photometry.
Indeed, \citet{MendezAbreu.etal.2017a} have recently used SDSS images to perform
a morphological analysis of CALIFA galaxies. The mass density maps, however, are
clearly of interest, despite the limitations in resolution. Bulge and disk
parameters deduced from an analysis of $\Sigma_\star$ images are not only more
physically meaningful, but also less ambiguous than those derived from the
surface brightness, which have the inconvenient property of depending on the
wavelength of the image.

\begin{figure}
\centering
\includegraphics{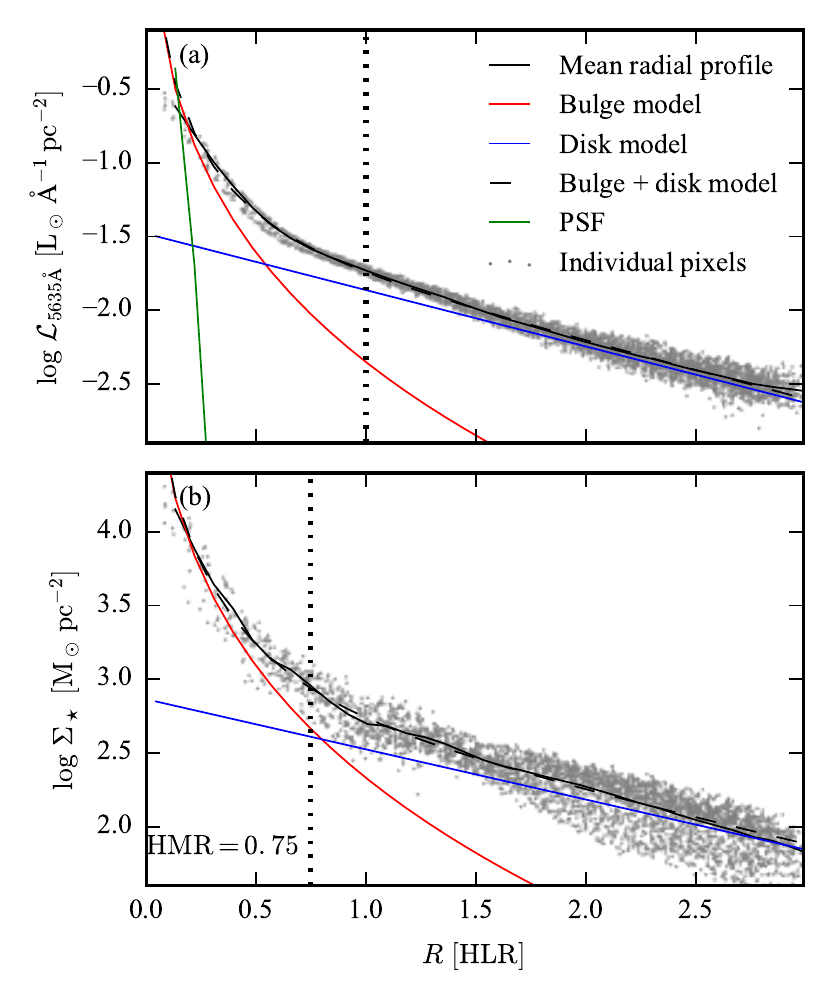}
\caption{
Bulge/disk fit of luminosity ({\em a}) and mass ({\em b}) surface
densities for CALIFA 0127. The morphology models (Equations
\ref{eq:disk-model} and \ref{eq:bulge-model}) were fitted to radial profiles of
these properties. The central $2.5\,"$ were masked to avoid issues with PSF
convolution (the green line shows the PSF profile). The vertical dotted line
marks $1\,\mathrm{HLR}$ in panel ({\em a}) and $1\,\mathrm{HMR}$ in panel ({\em
b}). Blue and red lines in both panels show respectively the fitted disk and
bulge models, while a dashed black line shows the combined models.}
\label{fig:morph-fit}
\end{figure}

Fig.\ \ref{fig:morph-fit} shows the radial profiles of
$\mathcal{L}_{5635\text{\AA}}$ (top) and  $\Sigma_\star$ (bottom) for the S0
galaxy CALIFA 127 (NGC 1349). The stellar mass density image is dezonified using
$\alpha = 1.2$, following the prescription outlined in Section
\ref{sec:dezonification}. Maps of physical properties for this galaxy are shown
in the appendix. Values for individual spaxels are shown as dots, while the
solid black line shows the azimuthally averaged profiles. Red and blue lines
show the results of a simplified bulge (red) + disk (blue) decomposition applied
to these 1D profiles. The bulge was modeled as following a classical de
Vaucouleurs $R^{1/4}$ profile:
\begin{equation}
\label{eq:bulge-model}
I_\mathrm{B}(R) = I_e\,\exp\left\{-7.669 \left[ \left( \frac{R}{R_e}
\right)^\frac{1}{4} - 1\right]\right\},
\end{equation}

\noindent while the disk was modeled as 
\begin{equation}
\label{eq:disk-model}
I_\mathrm{D}(R) = I_0\,\exp\left( -\frac{R}{h} \right).
\end{equation}

\noindent The parameters in these expressions have their usual meaning: $R_e$ is
the bulge effective radius, $I_e$ is the intensity at $R = R_e$, $h$ is the disk
scale radius, and $I_0$ the central intensity of the disk component. The total
profile is simply $I(R) = I_\mathrm{B}(R) + I_\mathrm{D}(R)$. The intensity $I$
refers to $\mathcal{L}_{5635\text{\AA}}$ in Fig.\ \ref{fig:morph-fit}a and to
$\Sigma_\star$ in Fig.\ \ref{fig:morph-fit}b. The $R < 2.5^{\prime\prime}$
region is excluded from the fit to minimise the effects of the PSF. A thorough
morphological analysis should be carried out in 2D and account for PSF smearing,
as well as consider a more general Sérsic function, but the crude approach
employed here suffices for our current purposes.

The bulge effective radii obtained from the fits are $R_e = 0.26\,\mathrm{HLR}$
for the $\mathcal{L}_{5635\text{\AA}}$ profile and $0.24\,\mathrm{HLR}$ for
$\Sigma_\star$ (both of the same order of the PSF width, and so not well
resolved), while the disk scales are $h = 1.14\,\mathrm{HLR}$ for
$\mathcal{L}_{5635\text{\AA}}$ and  $1.28\,\mathrm{HLR}$ for $\Sigma_\star$. It
thus seems that the structures have approximately the same sizes whether
measured in light or in mass. The relative bulge-to-disk intensity scales,
however, change substantially from the $\mathcal{L}_{5635\text{\AA}}$ to the
$\Sigma_\star$ fits, as indeed visually evident comparing the vertical scales in
Figs.\ \ref{fig:morph-fit}a and b. The bulge-to-total ratios implied by the fits
are $B/T = 0.39$ in light and $0.55$ in mass. In other words, the bulge is more
relevant in mass than in light, as expected from its older stellar populations
(see Fig.\ \ref{fig:K0127maps-physical}) and thus larger mass-to-light ratios.

These differences imply that this galaxy is more compact in mass than in light.
A convenient and simple way of quantifying this effect is by comparing the half
light and half mass radii (HLR and HMR, respectively). The HMR can be computed
using $\Sigma_\star$, finding the distance where the cumulative sum of the mass
reaches half of its total value.
For CALIFA 0127 we obtain $\mathrm{HLR} = 11.4^{\prime\prime}$ and $\mathrm{HMR}
= 8.5^{\prime\prime}$, as indicated by the dashed vertical lines in Fig.\
\ref{fig:morph-fit}. This galaxy is thus some $25\%$ more compact in mass than
in light, as indeed found for most CALIFA galaxies
\citep{GonzalezDelgado.etal.2015b}.

Interestingly, Fig.\ \ref{fig:morph-fit}b shows that the bulge and disk
components have almost exactly the same $\Sigma_\star$ at $R = 1\,\mathrm{HMR}$.
Thus, at least for this galaxy, $R = 1\,\mathrm{HMR} = 0.75\,\mathrm{HLR}$ is a
good place to divide the galaxy into bulge and disk-dominated regions. Previous
work by  our group complies with this recipe, treating regions within
$0.5\,\mathrm{HLR}$ as bulge-dominated and those at $R > 1\,\mathrm{HLR}$ as
disk-dominated \citep{GonzalezDelgado.etal.2014a}.

\subsection{Examples of usage for many galaxies}
\label{sec:example-many}
While there is much to be learned examining a single galaxy from the sample, the
true power of the catalogue becomes evident when we combine large chunks of it.
To illustrate this, we begin by investigating half mass and half light radii for
all galaxies in the sample. Then we move to a little processing, combining
radial profiles of all galaxies and deriving a mass--metallicity relation. We
finish this section obtaining a Schmidt-Kennicutt-like relation for a subsample
of spiral galaxies.

\subsubsection{Half mass vs. half light radii}
As discussed at the end of Section \ref{sec:morph-fit}, stellar mass and light
are not equality distributed spatially. \citet{GonzalezDelgado.etal.2015b} have
previously compared the HLR and HMR radii of 312 CALIFA galaxies.

\begin{figure}
\centering
\includegraphics{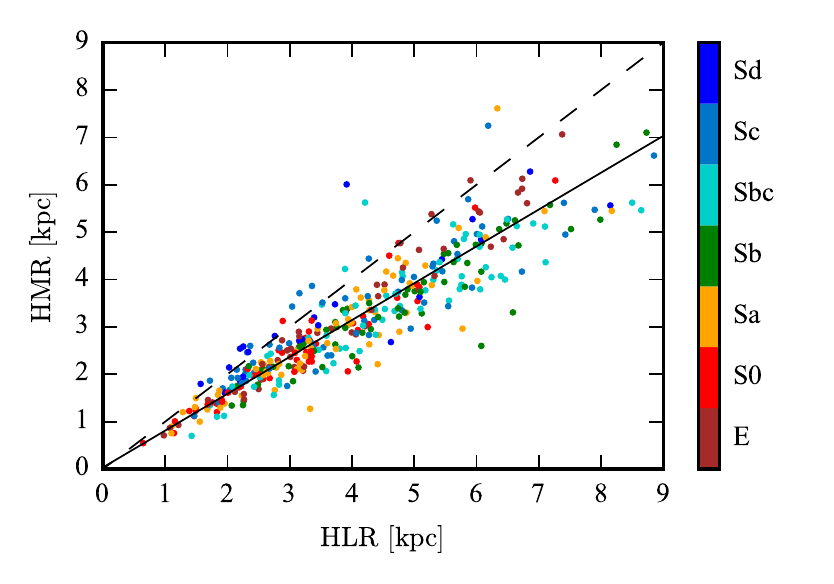}
\caption{Half Mass Radius (HMR) {\em versus} Half Light Radius
(HLR) for all galaxies in the sample. The $y=x$ line is shown as a black dashed
line, while the best linear fit is shown as a continuous black line.The color
of the points denotes the morphological type.} 
\label{fig:HLRxHMR-relation}
\end{figure}

Fig.\ \ref{fig:HLRxHMR-relation} updates that previous result, now for all 395
galaxies in our database, and with the slight changes in the reduction pipeline
and \starlight analysis described in Section \ref{sec:analysis}. The plot shows
HMR and HLR for each galaxy in the sample. For most galaxies, $\mathrm{HMR} <
\mathrm{HLR}$. A linear fit is drawn as a black line.
We find that the galaxies are on average $22\%$ more compact in mass. The
difference goes from $18\%$ for elliptical galaxies up to $25\%$ for Sbc
galaxies, and decreasing again to $7\%$ for Sd galaxies, in agreement with the
results reported by \citet{GonzalezDelgado.etal.2015b} for a smaller sample.

\subsubsection{Local relations: stellar mass surface density, age, and
metallicity}
\label{sec:mz-relation}
Galaxies come in various shapes and sizes. If we wish to combine and analyse all
the galaxies in the sample, we have to somehow homogenise the data. One such way
is to take mean radial profiles of the maps, using the HLR as a spatial scale.
We use radial bins of $0.1\,\mathrm{HLR}$, from $0$ to $2.5\,\mathrm{HLR}$.
As seen in Section \ref{sec:radprof}, we also take the ellipticity and
orientation into account.

\begin{figure}
\centering
\includegraphics{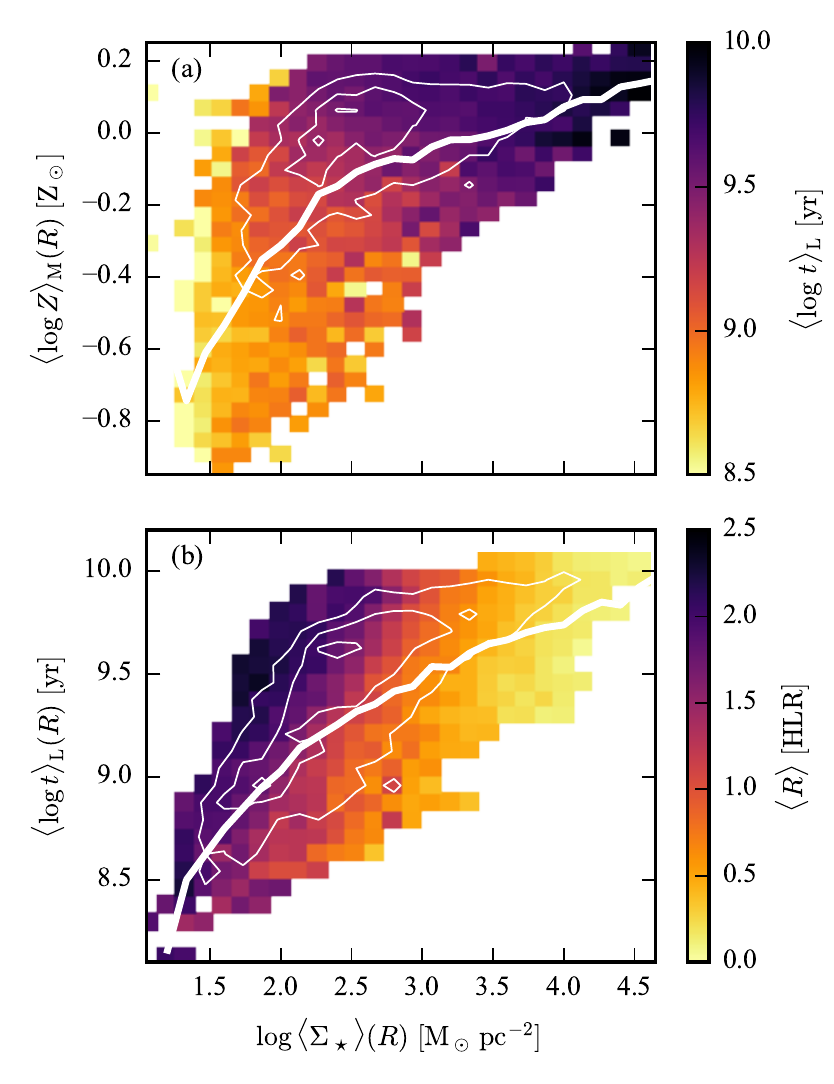}
\caption{{\em Panel (a)} Mass--metallicity relation using radial
profiles of $\langle \log\ Z \rangle_\mathrm{M}$ and $\Sigma_\star$, for all
galaxies in the sample. The radial bins cover $0$ to $2.5\,\mathrm{HLR}$ in
steps of $0.1\,\mathrm{HLR}$. The background tiles are the mean stellar age in
2D bins.
{\em Panel (b)} The same as panel (a), but for $\langle \log\,t
\rangle_\mathrm{L}$ as a function of $\Sigma_\star$, with colors representing
the mean for the mean radial distance.
The contours of the density of points is shown in thin white lines, and the mean
value of stellar metallicity and age, in bins of mass surface density, is shown
as a thick white line, in both panels.}
\label{fig:MZ-relation}
\end{figure}

By using the radial profiles of the maps $\Sigma_\star$ and $\langle \log\ Z
\rangle_\mathrm{M}$, we can build a mass--metallicity relation (MZR), as seen in
Fig.\ \ref{fig:MZ-relation}a. The color of the bins represents the mean value of
$\langle \log\,t \rangle_\mathrm{L}$. This plot reproduces one of the results
presented by \citet[figure 2]{GonzalezDelgado.etal.2014a}. There we used radial
profiles of 300 galaxies. Here we update that result for 395 galaxies from the
main sample.

In the bottom panel we show a similar plot, using $\Sigma_\star$ and $\langle
\log\,t \rangle_\mathrm{L}$, with colors representing the radial distance in HLR
units. The white line shows the mean value of stellar metallicity (top) and age
(bottom) in bins of mass surface density. The region where the MZR is steeper
corresponds to the outer parts of the galaxies, which also tend to be less dense
and have younger stellar populations. The MZR becomes flatter in the inner and
more dense parts, where the stellar populations are older.

\subsubsection{The $\Sigma_{\mathrm{SFR}}$--$\tau_V$ relation: A proxy for the SK law}
\label{sec:sk-law}
One of the most important relations in galaxy studies is the so called
Schmidt-Kennicutt (SK) law, which relates $\Sigma_{\mathrm{SFR}}$ to surface gas
densities \citep{Kennicutt.Evans.2012a}. Our \starlight-based estimates of
$\Sigma_{\mathrm{SFR}}$ differ from conventional ones, like those based on the
H$\alpha$ luminosity or the far-infrared emission, but at least we have a direct
estimate of $\Sigma_{\mathrm{SFR}}$. The same cannot be said about the gas
surface density $\Sigma_{\mathrm{gas}}$, the $x$-axis in the SK law, which
requires measurements of the cold gaseous component in galaxies.

We do, however, have $\tau_V$ at hand. Interpreting it as a proper dust optical
depth and making some assumption about grain sizes one may derive dust column
densities. Further assuming a constant dust-to-gas ratio one has that $\tau_V
\propto \Sigma_{\mathrm{gas}}$. These are all bold and highly questionable
assumptions, of course. Nevertheless they prompt us to investigate the relation
between $\Sigma_{\mathrm{SFR}}$ and $\tau_V$.

\begin{figure}
\centering
\includegraphics{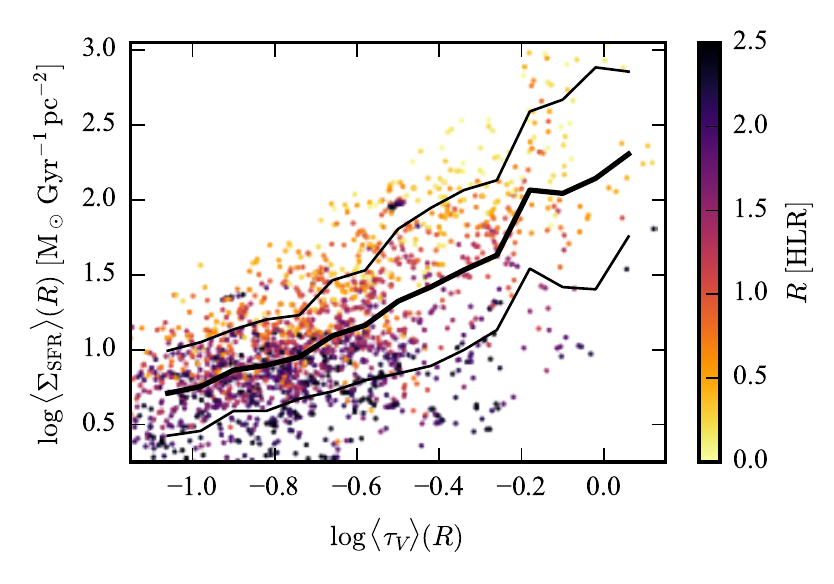}
\caption{Star formation rate surface density {\em versus} the dust extinction
(both obtained from our \starlight\ analysis) for $98$ face-on spiral galaxies,
as derived from images provided in our database. The points are radial means,
taken in bins of $0.1\,\mathrm{HLR}$, up to $2.5\,\mathrm{HLR}$. The color of
the points codes for the distance of the bin to the center of the galaxy, in
units of HLR. The mean value of $\log\,\langle \Sigma_\mathrm{SFR} \rangle$ in
bins of $\log\,\langle \tau_V \rangle$ is shown as a thick black line, with
$1\sigma$ range drawn as thin lines. The slope of this line, $\sim 1.3$, is
proxy for a Schmidt-Kennicutt relation.}
\label{fig:SK-like}
\end{figure}

In order to do this we first select a subsample of 98 galaxies classified as
spiral, non-interacting, with an ellipticity $< 0.2$ -- that is, close to being
face-on. These galaxies are processed in the same way as the previous section,
by calculating radial profiles of $\Sigma_{\mathrm{SFR}}$ and $\tau_V$. Fig.\
\ref{fig:SK-like} shows the results, with the color of the points encoding the
radial distance (in HLR units). The black line shows the mean value of mean star
formation rate in bins of $\tau_V$, with a $\pm 1 \sigma$ range drawn as thin
black lines. If we fit a line to the points, we obtain a slope of $\sim 1.3$.
This is amazingly close to the slope of the SK relation
\citep{Kennicutt.Evans.2012a}.
A preliminary version of this same relation was
presented in \citet[figure 3]{CidFernandes.etal.2015a}.

\section{Combining data from other sources}
\label{sec:example-other}
The combination of our \pycasso maps with data from other sources should be
useful both as an external test on the consistency of the properties offered in
our catalogue, and as a way to gain insight into physical processes within
galaxies. In this section we explore both approaches. We start by comparing the
stellar surface mass densities presented here with those obtained from near-IR
imaging (Section \ref{sec:S4G}). We then combine our stellar population
properties with emission line data to test our SFR estimates and to exemplify
how to explore the complementarity of these independent sources of information
(Section \ref{sec:Pipe3D}). In both cases only publicly available data are used.

\subsection{Mass maps from \sfourg}
\label{sec:S4G}
The \textit{Spitzer} Survey of Stellar Structure in Galaxies
\citep[\sfourg]{Sheth.etal.2010a} produced, among other results, maps of the
mass in old stars, through imaging in the $3.6$ and $4.5\,\mu\mathrm{m}$ bands.
We found 35 galaxies present in both \pycasso and \sfourg catalogues. In order
to compare their $\Sigma_\star$ maps to ours we first downloaded their dust-free
surface brightness maps at $3.6\,\mu\mathrm{m}$ ($S_{3.6\mu\mathrm{m}}$). As
prescribed by \citet{Querejeta.etal.2015a}, a mass-to-light ratio
$M/L_{3.6\mu\mathrm{m}} = 0.6\,\mathrm{M}_\odot\mathrm{L}^{-1}_\odot$ was used
to transform $S_{3.6\mu\mathrm{m}}$ to stellar surface mass densities. The
images were then aligned to ours assuming the centroids in the infrared and
optical are at the same position, and radial profiles
$\Sigma_\star^\mathrm{S^4G} (R)$ were computed exactly as described in Section
\ref{sec:radprof}.

\begin{figure}
\centering
\includegraphics{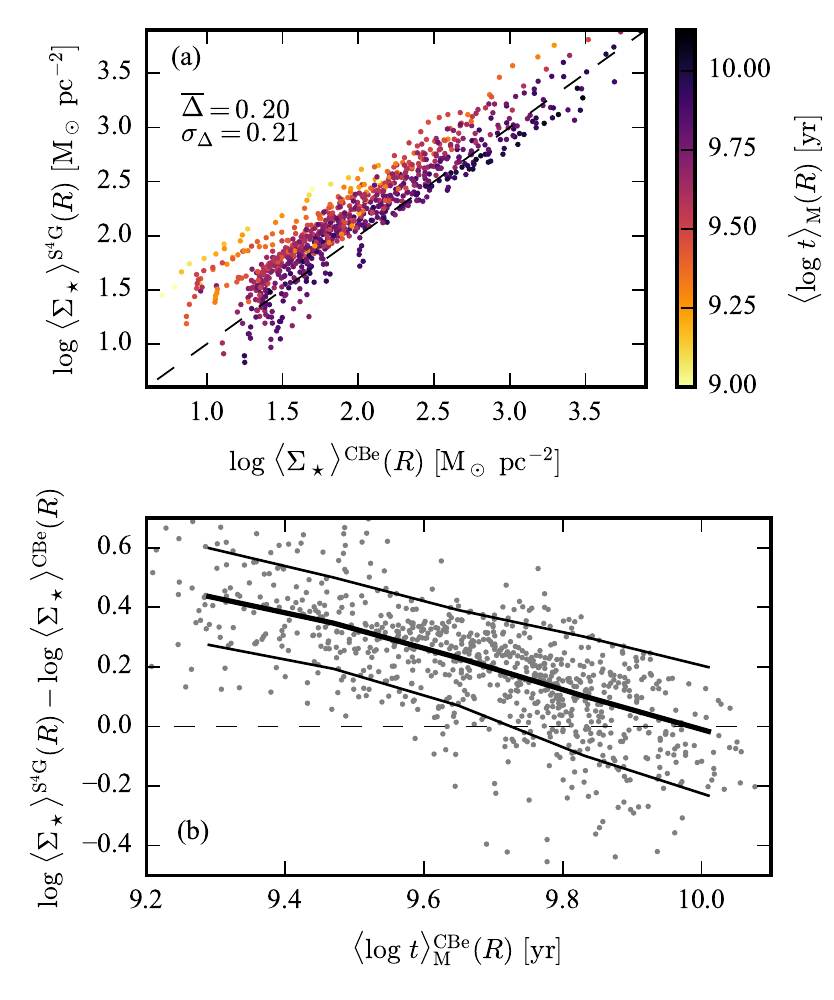}
\caption{Comparison between stellar mass surface density of 35 galaxies
from \pycasso\ (using base {\em CBe}) and \sfourg. Each point is the average in
an elliptical annulus around the centre of the galaxy. {\em  (a)} \sfourg\ {\em
versus} \pycasso. The colour of the points indicates the mass-weighted mean
stellar age, $\langle \log\,t \rangle_\mathrm{M}$. {\em  (b)} Difference between
the (logarithmic) masses in both samples, as a function of mass-weighted mean
stellar age. The thick black line shows the median in bins of age, and the thin
lines indicate the $\pm 1\sigma$ dispersion.}
\label{fig:s4g_mass}
\end{figure}

The comparison between the mass surface density from our catalogue and \sfourg
is shown in Fig.~\ref{fig:s4g_mass}. Because \sfourg assumes a Chabrier IMF, we
use the {\em CBe} base maps. On average these completely independent estimates
of $\Sigma_\star$ differ by $0.20$ dex, with an rms of $0.21$ dex, compatible
with the expected uncertainties of $0.08$ and $0.1$ dex from \starlight and
\sfourg, respectively.

There is, however, a clear trend for the difference to increase as
$\Sigma_\star$ decrease. The mass-weighted mean stellar age ($\langle \log\,t
\rangle_\mathrm{M}$), coded by the colours in Fig.~\ref{fig:s4g_mass}a, help us
understand what is behind this trend. As shown in Fig.~\ref{fig:s4g_mass}b, the
\sfourg and \pycasso masses are essentially identical as when $\sim 10^{10}$ yr
populations dominate the mass in stars. As the contribution of younger
populations increases the two estimates start to diverge, reaching a difference
of $\sim 0.4$ dex in the extreme cases. The excellent agreement in
$\Sigma_\star$ for large values of $\langle \log\,t \rangle_\mathrm{M}$ is
expected, as \citet{Querejeta.etal.2015a} assumes the light in
$3.6\,\mu\mathrm{m}$ comes from old populations, a hypothesis which is encoded
in the constant value adopted for $M/L_{3.6\mu\mathrm{m}}$. The offset for
younger regions is not much larger than the expected uncertainties in the
$\Sigma_\star$ estimates, and can be improved with a small age-dependent
correction to $M/L_{3.6\mu\mathrm{m}}$.

\subsection{Emission lines from Pipe3D}
\label{sec:Pipe3D}
Pipe3D \citep{Sanchez.etal.2016b} is a pipeline that derives stellar and ionised
gas properties from IFS. It uses FIT3D \citep{Sanchez.etal.2016a}, a spectral
synthesis tool, to model the stellar continuum. The results of Pipe3D applied to
200 galaxies (V500 setup) from CALIFA DR2 \citep{GarciaBenito.etal.2015a} are
publicly available at the CALIFA web site\footnote{CALIFA science dataproducts
available at
\url{http://www.caha.es/CALIFA/public_html/?q=content/science-dataproducts}.}.
We combine the emission line results from Pipe3D to \pycasso stellar data to
explore the gas properties in relation to those of the underlying stellar
populations.

When considering a combination of \pycasso and Pipe3D data, some issues should
be noted. The data reduction changed from CALIFA DR2 to DR3. The cubes are not
spatially compatible. The techniques used in Pipe3D were implemented
independently of this work. Also, Pipe3D used V500 cubes, while we used COMBO
cubes. All these characteristics mean that, while using data from the same
survey, Pipe3D and \pycasso are datasets different enough so that a comparison
between them is not trivial.

To match the cubes from both sources, we take a $58\times58$ pixels slice around
the reference pixel in each FITS file, taking advantage of the same pixel scale
and orientation in both cases. This allows us to compare results pixel by
pixel\footnote{Pipe3D used Voronoi binning in its analysis, but the actual
location of the zones, as well as the dezonification scheme, is different from
the ones used in \pycasso.}, as well as using radial bins. We exclude galaxies
that are morphologically irregular or interacting to avoid problems with the
centroid in the reference pixel. The final selection contains 166 galaxies
present in both samples.

\subsubsection{Star formation rates}
As a first application, we may obtain the SFR from the maps of
$\mathrm{H}\alpha$ emission. This is a measurement of SFR using a region of
spectra that is ignored by \starlight, being a totally independent measurement.
It is therefore instructive to compare these two estimates, following
\citet{Asari.etal.2007a}. This is the first time this type of study is done for
spatially resolved data.

The SFR is directly converted from  $\mathrm{H}\alpha$  luminosity using the
standard linear relation from \citet{Kennicutt.1998a}, the same used by
\citet{Sanchez.etal.2016b} to compare their results to other works. The
$\mathrm{H}\alpha$ fluxes are first corrected for dust attenuation using the
\citet*{Cardelli.Clayton.Mathis.1989a} law, with $R_V = 3.1$.
In this process, we assume an intrinsic ratio $\mathrm{H}\alpha /
\mathrm{H}\beta = 2.86$. We mask pixels where $\mathrm{S}/\mathrm{N} < 5$ in
$\mathrm{H}\alpha$ or $\mathrm{H}\beta$. Pixels where $W_{\mathrm{H}\alpha} <
14\,\text{\AA}$ are also masked, to ensure the $\mathrm{H}\alpha$ emission is
dominated by star formation (Lacerda et al., in prep.). The dereddened
$\mathrm{H}\alpha$ surface brightness translates straightforwardly to
$\Sigma_{\mathrm{SFR},\mathrm{H}\alpha}$, which is then averaged in radial bins
of $0.1\,\mathrm{HLR}$, between $0$ and $2.5\,\mathrm{HLR}$.

\begin{figure}
\centering
\includegraphics{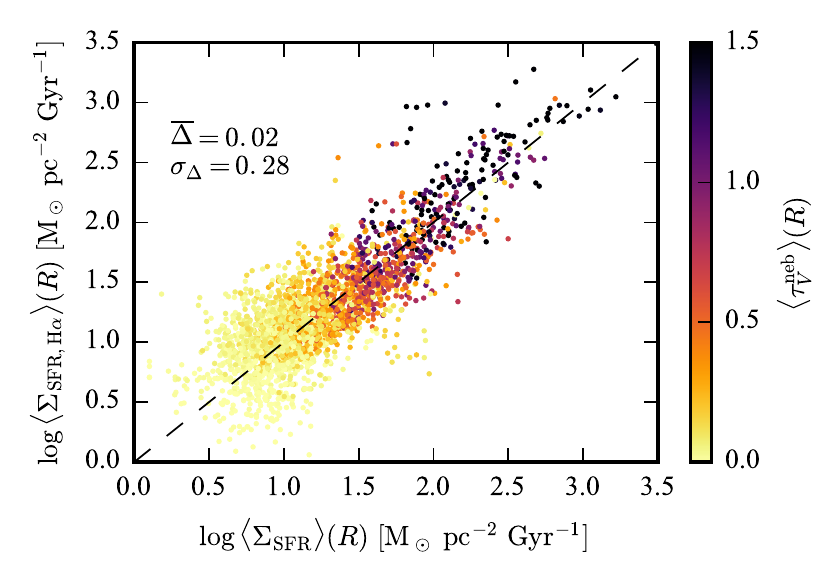}
\caption{Comparison between the star formation rate
surface density from 166 galaxies in \pycasso and
Pipe3D catalogues. Each point represents an elliptical annulus around the
nucleus, going from $0$ to $2.5\,\mathrm{HLR}$ in steps of
$0.1\,\mathrm{HLR}$. The points are coloured according to the mean nebular
attenuation.}
\label{fig:pipe3d-sfr}
\end{figure}

Fig.~\ref{fig:pipe3d-sfr} compares $\Sigma_{\mathrm{SFR},\mathrm{H}\alpha}$ with
our \starlight-based $\Sigma_\mathrm{SFR}$. The values from both sources agree
remarkably well, with virtually no offset.
The scatter in this plot, $0.28$ dex, is close to the estimated uncertainty of
$\Sigma_\mathrm{SFR}$ in Table \ref{tab:uncertainties}.
Points in Fig.~\ref{fig:pipe3d-sfr} are colour-coded by $\tau_V^\mathrm{neb}$,
the optical depth derived from the observed Balmer decrement.
We see that $\tau_V^\mathrm{neb}$ increases with increasing
$\Sigma_\mathrm{SFR}$, as expected from the pseudo Schmidt-Kennicutt law
previously investigated in Fig.\ \ref{fig:SK-like} in the context of stellar
properties alone. This remarkable agreement gives us confidence that the SFR
from the \pycasso database can be used not only for star-forming galaxies, but
for other types of galaxies where $\mathrm{H}\alpha$ is unrelated to star
formation.

\subsubsection{The WHAN diagram}
Still using Pipe3D data, we explore the classification scheme in the WHAN
diagram, which plots the equivalent width of H$\alpha$ ($W_{\mathrm{H}\alpha}$)
as a function of the ratio between the fluxes in the lines of
[N~\textsc{ii}]$\lambda 6584$ and H$\alpha$. This is a diagnostic diagram
proposed by \citet{CidFernandes.etal.2011a} to distinguish among the main
ionisation sources behind galaxy spectra. Star formation (SF) and active
galactic nuclei (AGN) are separated by vertical lines (constant
$[\mathrm{N~\textsc{ii}}]/\mathrm{H}\alpha$) in this diagram, while sources
located below $W_{\mathrm{H}\alpha} = 3\,\text{\AA}$ are associated to
photoionisation by hot low-mass evolved stars (HOLMES).

\begin{figure}
\centering
\includegraphics{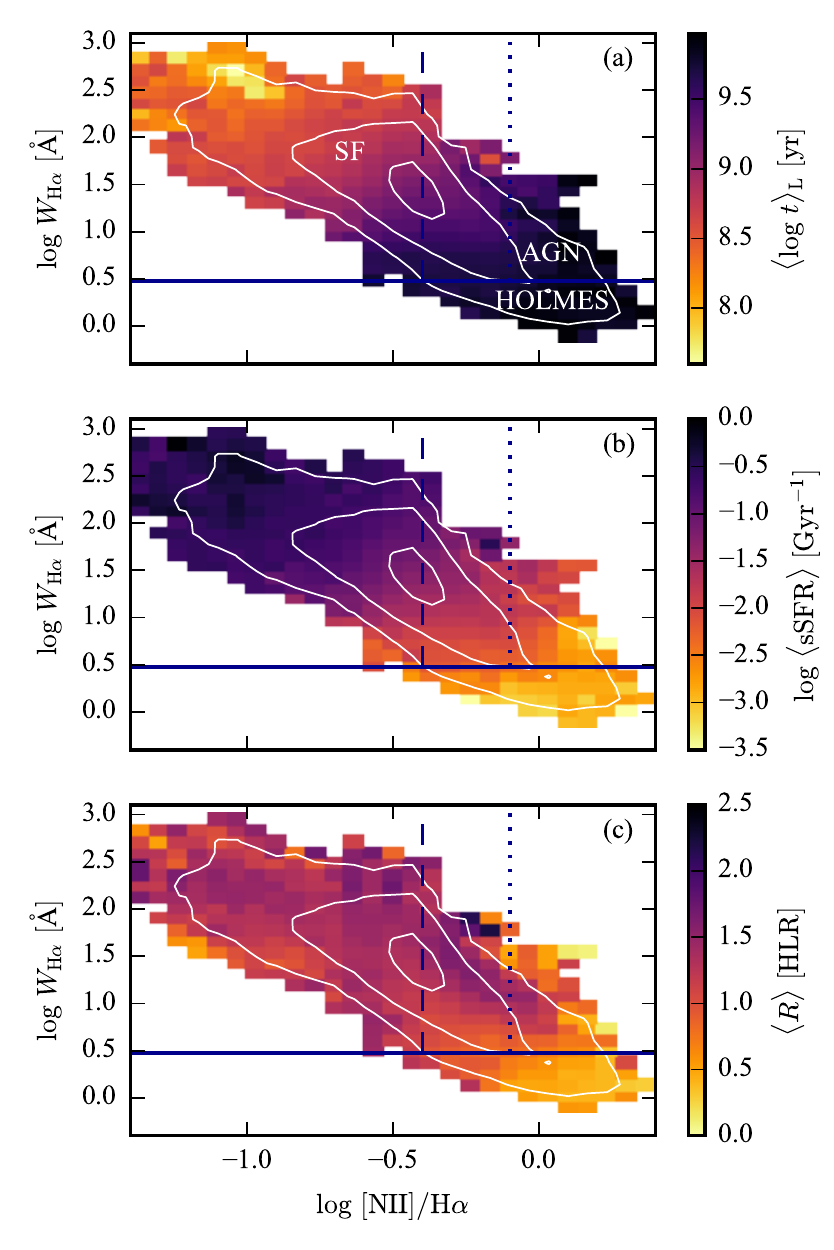}
\caption{Histogram of equivalent width of $\mathrm{H}\alpha$ versus
  the line ratio $[\mathrm{N~\textsc{ii}}]/\mathrm{H}\alpha$, known as
  WHAN diagram, using all pixels (no radial binning) from 166 galaxies
  matched to Pipe3D maps.  The colour of the bins represent the mean
  value of luminosity-weighted mean stellar age (panel a), specific star
  formation rate (panel b) and radial distance (panel c). Each bin has
  at least 5 data points.  White contours indicate the logarithm of the
  density of data points. Blue lines separate the WHAN diagram in the
  classes SF (star-forming), AGN and HOLMES. The horizontal continuous
  line shows the $W_{\mathrm{H}\alpha} = 3\,\text{\AA}$ limit for
  HOLMES. The dashed and dotted vertical lines are equivalent to the
  \citet{Stasinska.etal.2006a} and \citet{Kewley.etal.2001a} SF/AGN
  dividing lines, respectively.}
\label{fig:pipe3d-whan}
\end{figure}

We extrapolate the original intent of the WHAN diagram by plotting not whole
galaxies, but regions of galaxies. In Fig.~\ref{fig:pipe3d-whan} we have
2D-histograms of all pixels of the 166 galaxies matched to Pipe3D. Bin colours
in each panel represent the average luminosity-weighted stellar age ($\langle
\log\,t \rangle_\mathrm{L}$, panel a), specific star formation rate
($\mathrm{sSFR}$, panel b) and distance to the nucleus in units of HLR (panel
c); only bins with 5 points or more are plotted. Contours show the logarithm of
the density of points. The $W_{\mathrm{H}\alpha} = 3\,\text{\AA}$ line
separating HOLMES-ionised gas (HIG) from SF and AGN regions is indicated. Lines
at $[\mathrm{N~\textsc{ii}}]/\mathrm{H}\alpha = -0.4$ and $-0.1$ separate SF
from AGN regions according to the criteria of \cite{Stasinska.etal.2006a} and
\cite{Kewley.etal.2001a}, respectively, as explained in
\citet{CidFernandes.etal.2010a}.

In Fig.~\ref{fig:pipe3d-whan}a we can see that the HOLMES-dominated regions in
the WHAN diagram is composed of old stellar populations. Those regions also have
the lowest sSFR, as seen in Fig.~\ref{fig:pipe3d-whan}b. This might not be
surprising, since $W_{\mathrm{H}\alpha}$ is a proxy for sSFR. This is only true,
however,  for systems where $\mathrm{H}\alpha$ is produced by H~{\scshape ii}
regions. In contrast, the sSFR measured from the stellar continuum can be used
to estimate star formation rates in systems where $\mathrm{H}\alpha$ is
dominated by radiation fields other than young stars, such as AGN and HOLMES.
Fig.~\ref{fig:pipe3d-whan}c shows that the region in the WHAN diagram dubbed AGN
is populated not only by the nuclear parts of galaxies, but also by off-nuclear
regions. We expect those at low $R$ to be {\em bona fide} AGN, while at higher
$R$ the ionisation field is most likely a mixture of H~{\scshape ii} regions and
HOLMES (see Lacerda et al., in prep).

\section{Conclusions}
\label{sec:conclusions}
We have presented a value-added catalogue, publicly available at
\url{http://pycasso.ufsc.br/}, for CALIFA galaxies analysed with the spectral
synthesis code \starlight and the \pycasso pipeline. The CALIFA IFS sample is
representative of galaxies in the local universe at $0.005 < z < 0.03$. Our
catalogue is comprised of 445 galaxies from the CALIFA Data Release 3 with COMBO
data. From these, 395 are taken from the mother sample, the remaining are from
the extended sample.

Our public data contains maps of stellar mass surface densities, mean stellar
ages and metallicities, stellar dust attenuation, star formation rates, and
kinematics. We have provided a road map on how to read and interpret each
physical-property image on our catalogue, and given a few examples of how to
access the data in the electronic files. Since our analysis is based on spectra
that have been binned with Voronoi tesselation to increase the signal-to-noise
ratio, we also discuss how one may dezonify (or pixelize) the information on
Voronoi zones.

We have also shown how to obtain radial profiles from our maps. Radial profiles
are useful representations of two-dimensional maps, and we showcase their
usefulness by separating a galaxy's bulge and disk both in its light and stellar
mass profiles. This case study agrees with our previous results
\citep{GonzalezDelgado.etal.2015b} that galaxies are more compact in mass than
in light.

We have used our catalogue to update and revisit a few of our former results,
like the relation between half mass and half light radii, and local relations
between stellar mass surface density, age, and metallicity.  A few previously
unexplored results were also presented, like a pseudo Schmidt--Kennicutt
relation between the SFR surface density and the dust optical depth obtained
using only stellar continuum information.
 
Finally, in order to illustrate the potential of the catalogue when used in
conjunction with complementary information from other databases, we have
combined it with publicly available data from the \sfourg and Pipe3D databases.
Using emission line measurements for 166 galaxies from Pipe3D we compare the
specific star formation rate from the stellar continuum to that obtained from
$\mathrm{H}\alpha$, finding that the two agree to with a factor of 2. We also
investigate how galaxy pixels populate the WHAN diagram, which can distinguish
the ionisation by SF, AGN or HOLMES in full galaxies. We show that the HOLMES/SF
scenario is compatible with the stellar population ages and sSFR in the \pycasso
database. These example applications demonstrate the catalogue should be a
useful tool to gain insight into galaxy physics.

\section*{acknowledgements}
ALA, NVA, RCF and EADL acknowledge the support from the CAPES CsF--PVE project
88881.068116/2014-01.

CALIFA is the first legacy survey carried out at Calar Alto. The CALIFA
collaboration would like to thank the IAA-CSIC and MPIA-MPG as major partners of
the observatory, and CAHA itself, for the unique access to telescope time and
support in manpower and infrastructures.  We also thank the CAHA staff for the
dedication to this project.

We thank the support from the Spanish Ministerio de Econom\'\i a y
Competitividad, through projects AYA2016-77846-P, AYA2014-57490-P,
AYA2010-15081, and Junta de Andaluc\'\i a FQ1580. We also thank the Viabilidad,
Dise\~no, Acceso y Mejora funding program, ICTS-2009-10, for funding the data
acquisition of this project.
ALdA, EADL, and RCF thanks the hospitality of the IAA and the support of CNPq.
RGD acknowledges the support of CNPq.

We thank the referee for comments which substantially improved this manuscript.

This research made use of Astropy, a community-developed core Python package for
Astronomy \citep{Astropy.etal.2013a}.

\bibliographystyle{mnras}
\bibliography{biblio.bib}

\appendix

\section{Other galaxies}

\begin{figure*}
\centering
\includegraphics{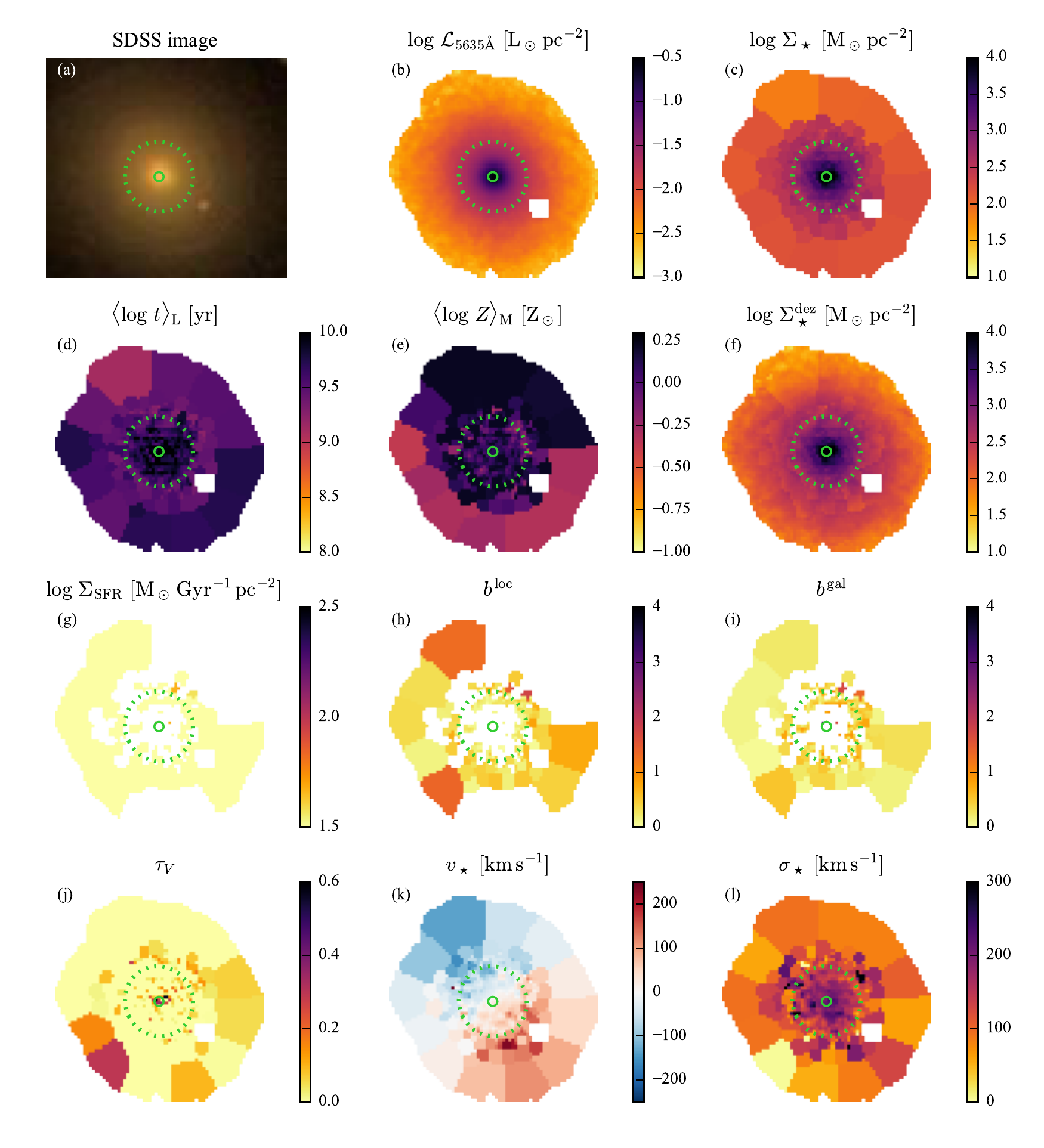}
\caption{Maps of selected physical properties for the elliptical galaxy NGC 1349
(CALIFA 0127). A dotted ellipse marks the $1\,\mathrm{HLR}$ area. }
\label{fig:K0127maps-physical}
\end{figure*}

\begin{figure*}
\centering
\includegraphics{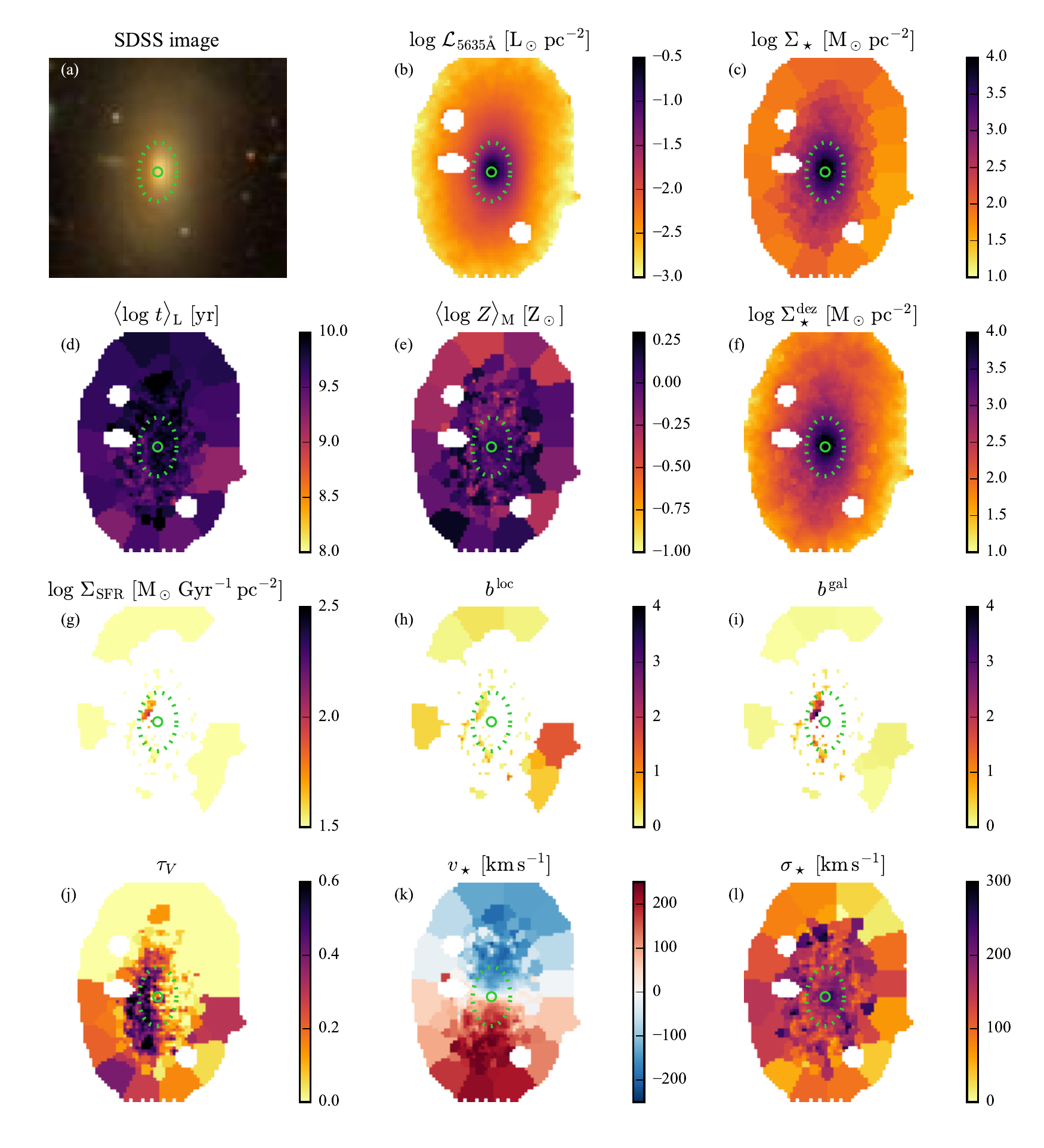}
\caption{As Fig.\ \ref{fig:K0127maps-physical}, but for the S0 galaxy UGC
10905 (CALIFA 0858).}
\label{fig:K0858maps-physical}
\end{figure*}

\begin{figure*}
\centering
\includegraphics{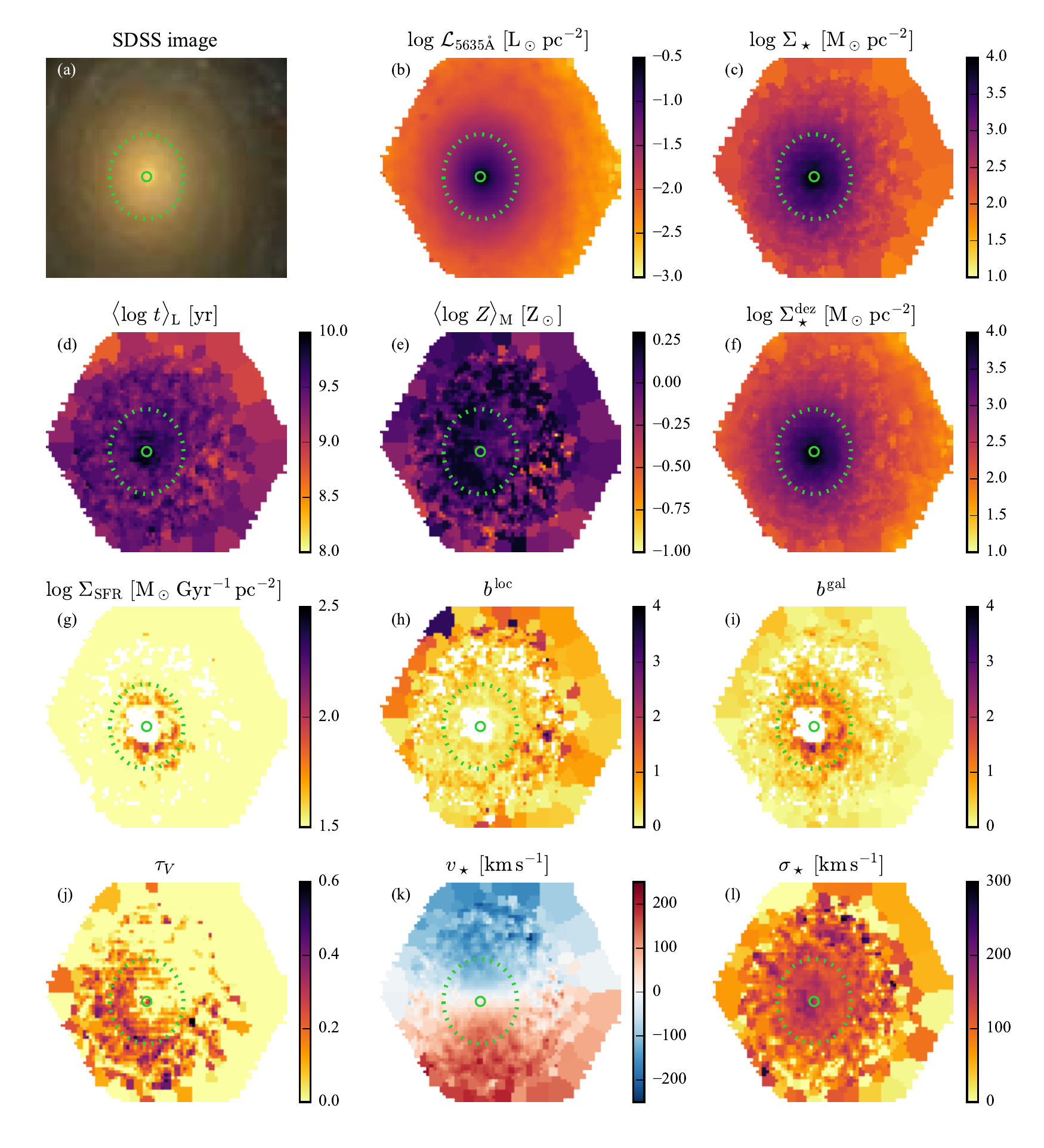}
\caption{As Fig.\ \ref{fig:K0127maps-physical}, but for the Sa galaxy NGC 1070
(CALIFA 0102).}
\label{fig:K0102maps-physical}
\end{figure*}

\begin{figure*}
\centering
\includegraphics{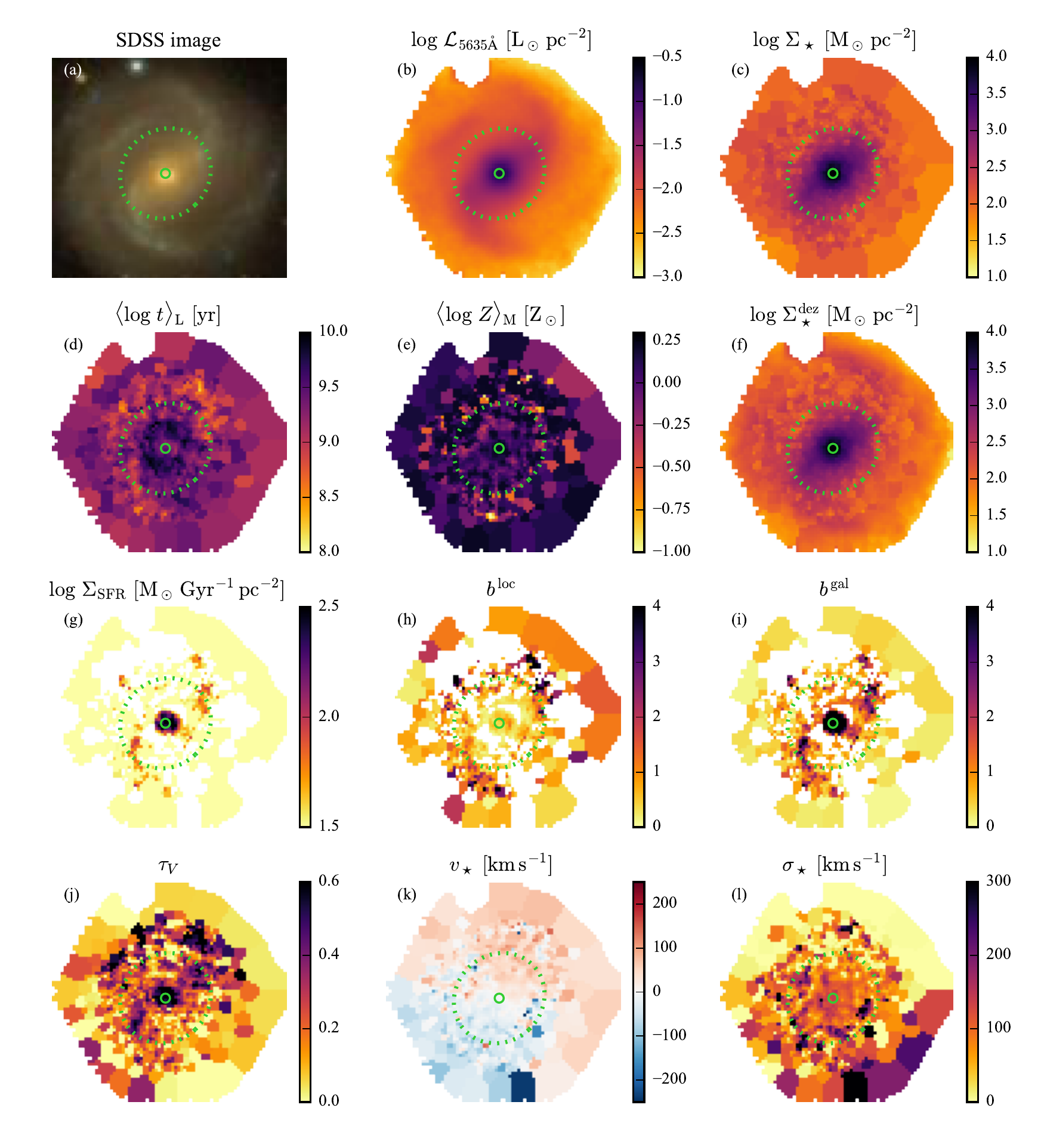}
\caption{As Fig.\ \ref{fig:K0127maps-physical}, but for the Sb galaxy NGC 0776
(CALIFA 0073).}
\label{fig:K0073maps-physical}
\end{figure*}

\begin{figure*}
\centering
\includegraphics{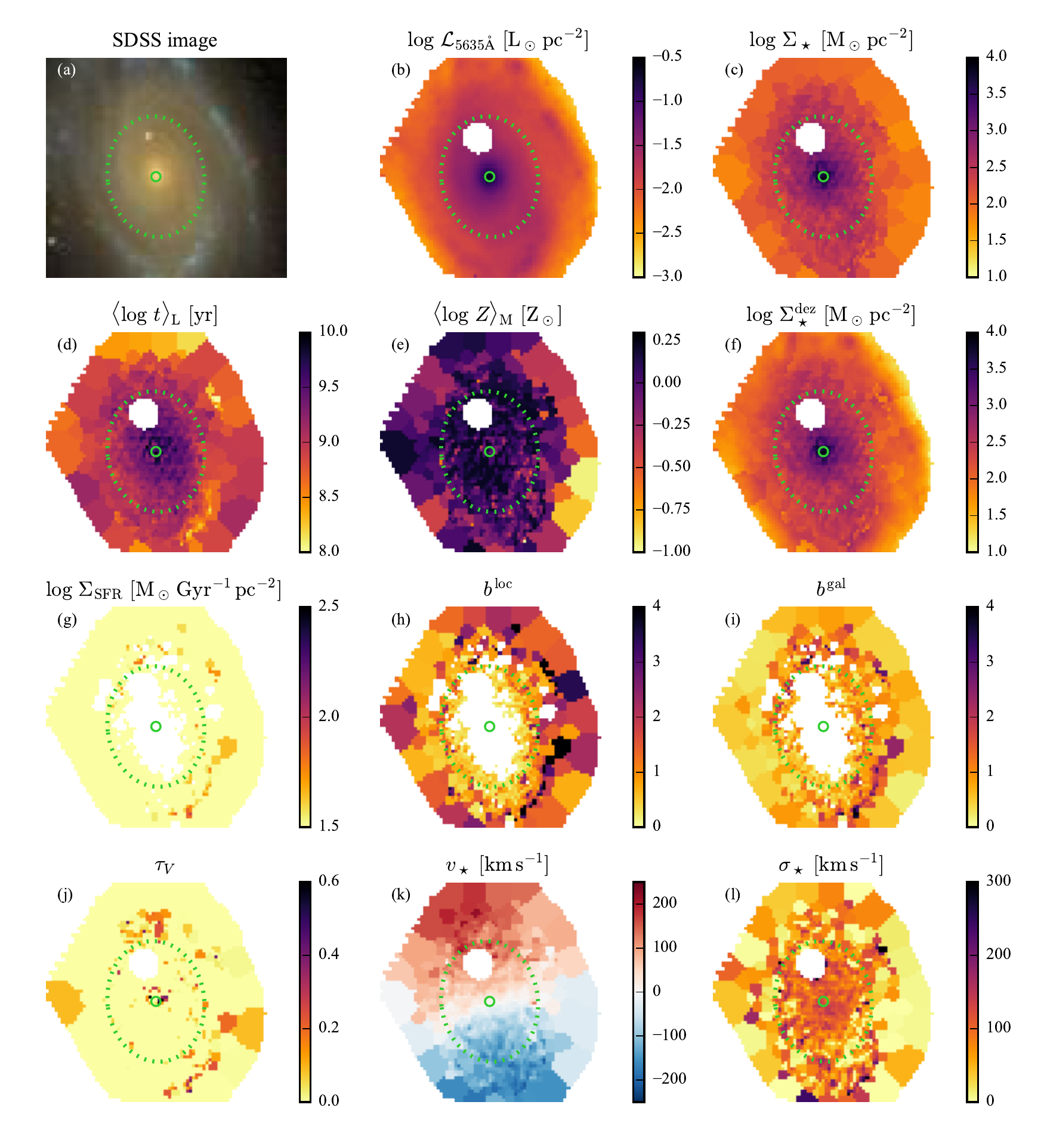}
\caption{As Fig.\ \ref{fig:K0277maps-physical}, but for the Sbc galaxy NGC
2916 (CALIFA 0277).}
\label{fig:K0277maps-physical}
\end{figure*}

\begin{figure*}
\centering
\includegraphics{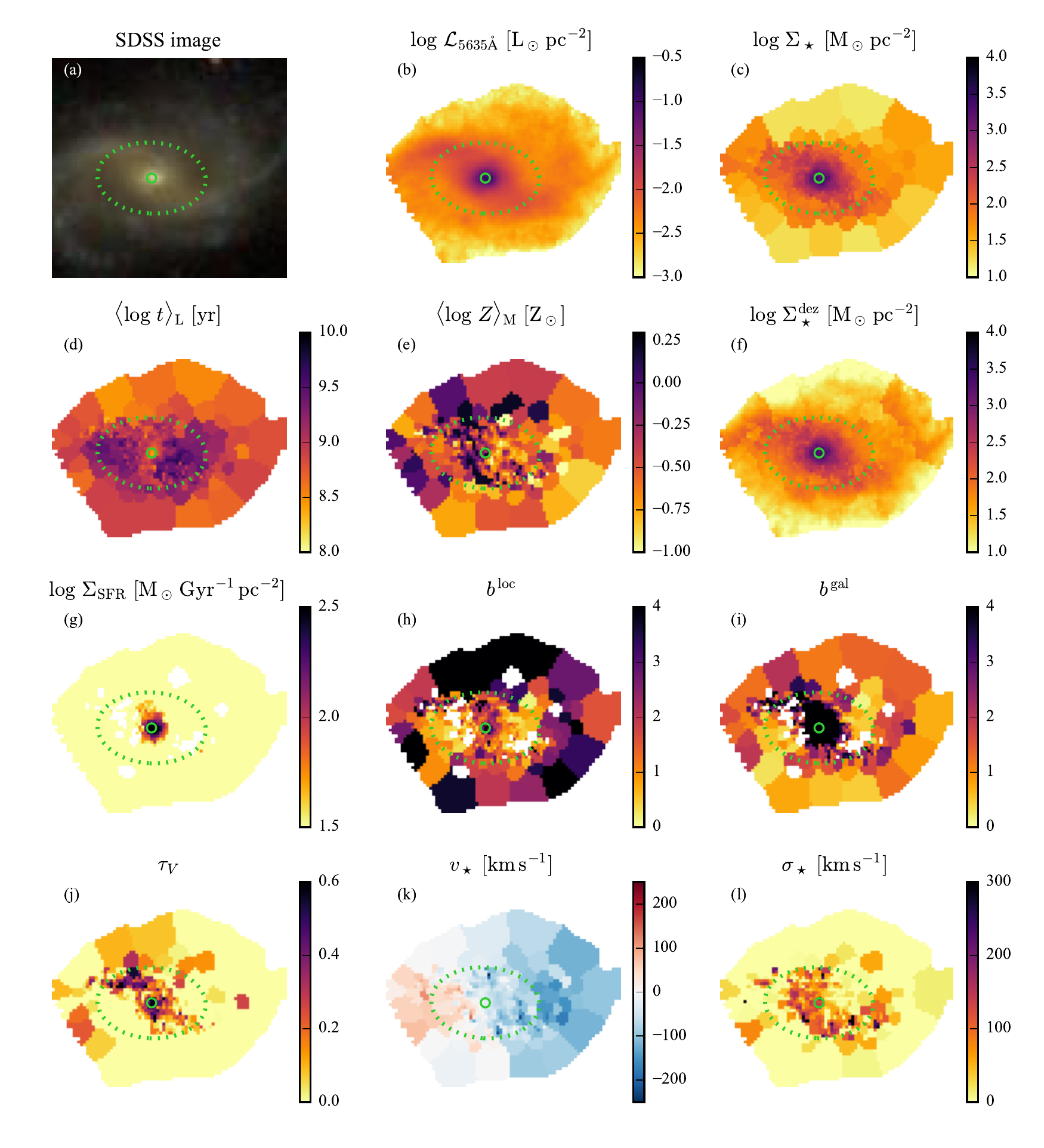}
\caption{As Fig.\ \ref{fig:K0127maps-physical}, but for the Sc galaxy NGC 7819
(CALIFA 0003).}
\label{fig:K0003maps-physical}
\end{figure*}

\begin{figure*}
\centering
\includegraphics{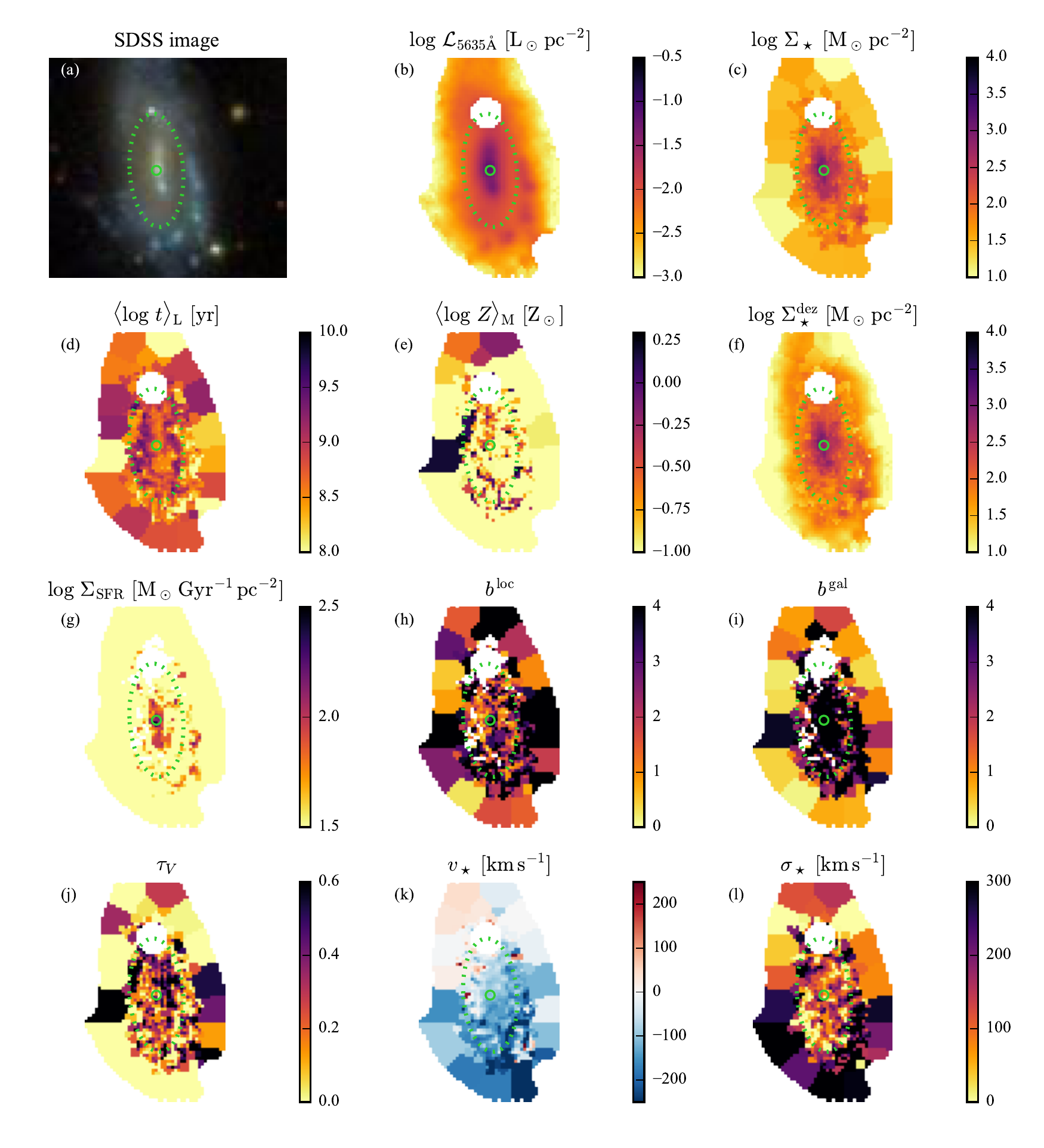}
\caption{As Fig.\ \ref{fig:K0127maps-physical}, but for the Sd galaxy UGC 00312
(CALIFA 0014).}
\label{fig:K0014maps-physical}
\end{figure*}

\begin{figure*}
\centering
\includegraphics{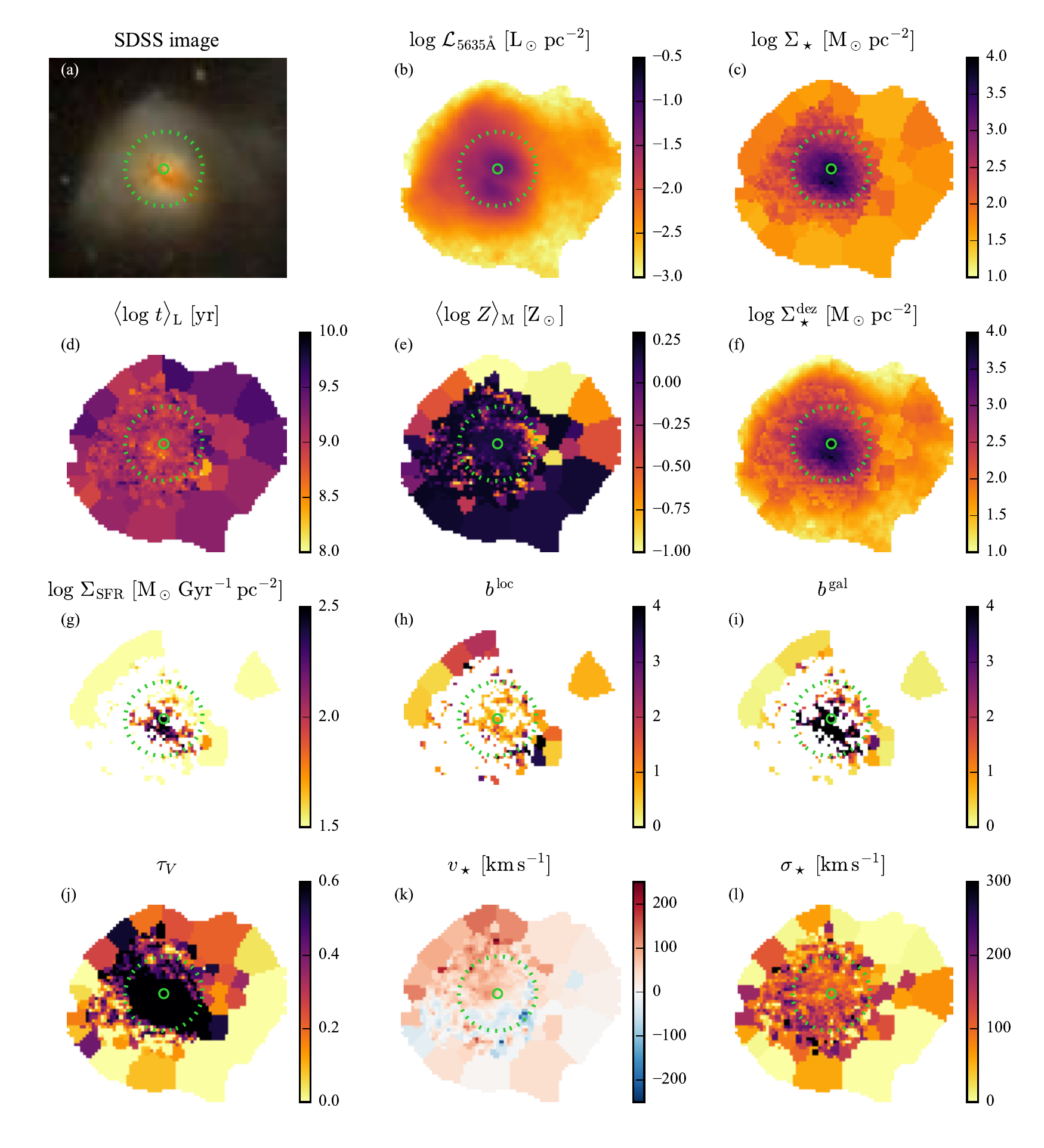}
\caption{As Fig.\ \ref{fig:K0127maps-physical}, but for the merger system Arp
220 (CALIFA 802).}
\label{fig:K0802maps-physical}
\end{figure*}

This appendix shows a few more examples of maps available in the \pycasso\
database (\url{http://pycasso.ufsc.br/}) presented in this paper.
Figs.\ \ref{fig:K0127maps-physical} to \ref{fig:K0802maps-physical} are sorted
according to morphological type, from an elliptical galaxy to a merger.

\bsp	
\label{lastpage}
\end{document}